\documentclass[aps,prb,twocolumn,amsmath,amssymb,superscriptaddress,scrartcl,eqsecnum,longbibliography,nofootinbib]{revtex4-2}

\usepackage{xcolor}
\usepackage{graphicx}
\usepackage{subfigure}
\usepackage{amsthm}
\usepackage{mathtools}
\usepackage{braket}

\usepackage{tikz}
\usetikzlibrary{decorations.pathmorphing}

\makeatletter
\PassOptionsToPackage{caption=false}{subfig} 
\usepackage{hyperref}
\hypersetup{
breaklinks=true,
colorlinks=true,
citecolor=blue,
linkcolor=black,
filecolor=black,
urlcolor=blue
}
\IfFileExists{lmodern.sty}{\usepackage{lmodern}}{}

\renewcommand{\bar}{\overline}

\renewcommand{\leq}{\leqslant}
\renewcommand{\geq}{\geqslant}

\newcommand{\dd}{\mathrm{d}}

\makeatletter

\newcommand*{\wideboxed}[1]{\setlength{\fboxsep}{1ex}%
  \fbox{\m@th$\displaystyle#1$}}
\makeatother

\def\be{\begin{equation}}
\def\ee{\end{equation}}

\makeatother

\begin{document}

\title{Detecting Higher Berry Phase via Boundary Scattering}

\author{Chih-Yu Lo}
\affiliation{School of Physics, Georgia Institute of Technology, Atlanta, GA 30332, USA}

\author{Xueda Wen}
\affiliation{School of Physics, Georgia Institute of Technology, Atlanta, GA 30332, USA}

\begin{abstract}

Higher Berry phase has recently been proposed to study  the topology of the space of gapped many-body quantum systems. In this work, we develop a boundary-scattering approach to detect higher Berry phases in one-dimensional gapped free-fermion systems. By coupling a gapless lead to the gapped system, we demonstrate that the higher Berry invariant can be obtained by studying the higher winding number of the boundary reflection matrix. The resulting topological invariant is robust against perturbations such as disorder. Our approach establishes a connection between higher Berry invariants and transport properties, thereby providing a potentially experimentally accessible probe of parametrized topological phases.

\end{abstract}
\maketitle



\section{Introduction}
\label{Sec:Introduction}

Higher Berry phase, the many-body generalization of Berry phase of parameterized quantum systems, was proposed 
to study the space of gapped many-body systems and, more broadly, the topological properties of parametrized gapped phases
\cite{kitaevSimonsCenter1,
kitaevSimonsCenter2,
kitaev2015talk,kitaev2019,
2010_Teo_Kane,
KS2020_higherberry, KS2020_higherthouless, Kapustin2201, 
Hsin_2020, Cordova_2020_i, Cordova_2020_ii, Else_2021,2022aBachmann,Choi_Ohmori_2022, 2023_Wen,Aasen_2022, 
	Hsin_2023,Shiozaki_2022, Ohyama_2022, ohyama2023discrete, Kapustin2305, homotopical2023, 2023Ryu, 2023_Qi,2023Shiozaki,2023Spodyneiko,2023Debray,
2024_Sommer1,2024_Sommer2,2024_Shuhei1,2024_Shuhei2,
2024_Multi_WF,2024_Geiko,
beaudry_2025,Kapustin_2025,choi_2025,wen_2025,2023_Prakash,2024_Prakashi,
Manjunath_2025,bose2025,2025_Jones,kubota2025,
Choi_2026,Brennan_2026,
Copetti_2025,Shiozaki_2025,2026_Else}.
This interest has been largely driven by Kitaev's conjecture \cite{kitaevSimonsCenter1,
kitaevSimonsCenter2,
kitaev2015talk}, which proposes that invertible phases of matter are classified by generalized cohomology theories.
To probe the topology of the space of gapped 
many-body systems, Kapustin and Spodyneiko proposed the concept of higher Berry curvatures for general gapped 
many-body lattice systems \cite{KS2020_higherberry,KS2020_higherthouless}.
This concept was subsequently developed in a systematic way within the framework of operator algebras, covering both symmetric and symmetry-breaking settings \cite{Kapustin2201}.
The flux of higher Berry curvature 
through a surface in the parameter space defines the higher Berry phase; when the surface is closed, this quantity becomes a topological invariant, often referred to as the higher Berry invariant.

\smallskip
Physically, a nonzero higher Berry invariant is associated with
a topological pump in a parametrized gapped system. More generally, the higher Berry curvature in $d$-dimensional systems can be interpreted as describing a flow of higher Berry curvature in $(d-1)$-dimensional systems \cite{2023_Wen}. In one dimension, this picture reduces to a flow of ordinary Berry curvature in real space, giving rise to the intriguing phenomenon of Chern number pumping \cite{2023_Wen,2023_Qi,2024_Sommer1,beaudry_2025}. 

Recent developments have clarified the wavefunction-level meaning of higher Berry invariants.
A nonzero higher-Berry invariant implies that the family of ground states carries a gerbe structure, generalizing the line bundle familiar from ordinary Berry phases.
When this gerbe is nontrivial, there is an obstruction to representing the ground states by a globally continuous matrix-product-state tensor over parameter space \cite{2023Ryu,2023_Qi}. 

Subsequent works have shown that higher Berry curvatures can be constructed directly from MPS representations, both in one dimension \cite{2024_Sommer1,2024_Shuhei1} and in higher dimensions \cite{2024_Sommer2,2024_Shuhei2}, providing a natural extension of the original Hamiltonian-based construction \cite{KS2020_higherberry}.

More recently, higher Berry phases have been extended to the study of spaces of conformal boundary conditions, or equivalently, conformal boundary states \cite{choi_2025,wen_2025}. One can define higher Berry connections and curvatures on the space of conformal boundary conditions using correlation functions of boundary-condition-changing operators \cite{choi_2025}. Physically, parametrized conformal boundary conditions can induce a flow of Berry curvature and, correspondingly, a Chern-number pump in boundary conformal field theories \cite{wen_2025}. This phenomenon closely mirrors the Berry-curvature flow in one-dimensional gapped systems \cite{2023_Wen,2023_Qi,2024_Sommer1,beaudry_2025}.

\smallskip

Higher Berry invariants also constrain phase diagrams \cite{Hsin_2020}.
If the parameter space $X$ lies entirely within a gapped region, the family of Hamiltonians defined over $X$
captures global topological information about that 
region.
From this viewpoint, conventional gapped phases correspond to connected components of the space of Hamiltonians, while parametrized families probe more refined topological structures.
This perspective has stimulated recent interest in parametrized systems and in the structure of critical points in phase diagrams \cite{Hsin_2020, 2023_Prakash, 2024_Prakashi, Manjunath_2025, bose2025, 2025_Jones, Shiozaki_2025, 2026_Else, 2026_Inamura_a, 2026_Inamura_b,2510_Yabo}.

\smallskip

Despite these conceptual advances, an important practical question remains open: How can higher Berry invariants be detected experimentally? More specifically, can information about these invariants be extracted solely from boundary probes, without accessing the bulk wavefunction or Hamiltonian directly?

\subsection{Motivations and Setup}

Our motivations are twofold.

\smallskip

First, from a conceptual perspective, it has recently been shown that coupling a conformal field theory to a parametrized gapped system drives the interface toward parametrized conformal boundary conditions under renormalization \cite{choi_2025,wen_2025}.
In this picture, the higher Berry phase of the bulk manifests as a Berry-curvature flow in the boundary CFT \cite{wen_2025}.
Since this flow is determined by how the boundary reflects gapless modes, it is natural to ask whether the higher Berry invariant can be extracted directly from the boundary data itself, without analyzing the induced spectral flow.

\smallskip

Second, from an experimental standpoint, directly measuring bulk higher Berry invariants is typically challenging.
If the same information is encoded in boundary scattering processes, then boundary probes may provide a practical route to detecting higher Berry phases.

 \smallskip

\begin{figure}[t]
\centering
\includegraphics[width=2.5in]{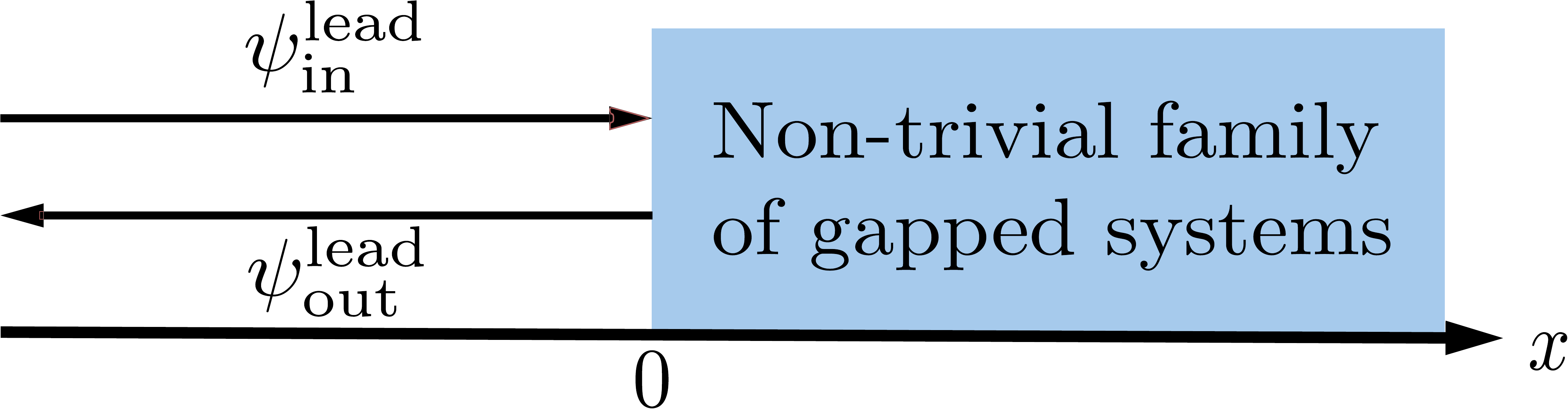}
\caption{Schematic illustration of detecting the 
topological invariant of a parametrized family of gapped systems via boundary scattering.
In the limit where the gapped system is infinitely long, an incoming wavefunction $\psi_{\mathrm{in}}$ from the left lead is completely reflected into an outgoing wavefunction $\psi_{\mathrm{out}}$. The resulting reflection matrix $R(\lambda)$ as defined in \eqref{def:Rmatrix} depends on the gapped-system parameters $\lambda$. As these parameters are varied, the evolution of $R$ in parameter space defines a quantized topological winding number.
}
\label{Fig:scheme}
\end{figure}

To address these questions, we consider a one-dimensional parametrized family of gapped systems coupled to a gapless lead, as illustrated in Fig.~\ref{Fig:scheme}.
The lead is in the region $(-\infty,0)$, while the parametrized gapped system resides in $(0,+\infty)$,
with the interface located at $x=0$.
Our goal is to show that the higher Berry invariant of the bulk family can be extracted entirely from the reflection matrix at this interface.

Throughout this work, we focus on free-fermion insulators with a global $U(1)$ charge symmetry. Although the presence of 
$U(1)$ symmetry is not strictly necessary for realizing higher Berry phases, it allows for a particularly transparent formulation in terms of charge transport and scattering. This perspective also facilitates experimental access, as the relevant quantities can, in principle, be probed through interference and electrical conductivity measurements \cite{2003_Ji}.

When the energy of an incoming electron from the lead lies within the bulk gap of the insulator, transmission is suppressed and the electron is completely reflected at the interface (See Fig.\ref{Fig:scheme})
\begin{equation}
\label{def:Rmatrix}
    \psi_\text{out}^\text{lead}(0) = R(\lambda) \,\psi_\text{in}^\text{lead}(0).
\end{equation}
That is, the total reflection is described by a unitary reflection matrix $R$, which depends smoothly on the parameters $\lambda$ characterizing the Hamiltonian of the gapped system.

We assume that the parameter space $X$ of the gapped Hamiltonians forms a smooth manifold. The reflection matrix $R$ then defines a smooth map
\be
\label{eq:map}
R: \,\, X\to U(N),
\ee
where $N$ in the unitary matrix $U(N)$ denotes the number of propagating channels in the lead.
For example, for a single spinless lead with one right- and one left-moving mode, we have $N=1$.

It is well known that, in the presence of global $U(1)$ symmetry, tuning the parameters along a closed loop in the parameter space
can realize a nontrivial Thouless charge pump \cite{thouless_1983}. 
The associated topological invariant can be expressed purely in terms of the reflection matrix as \cite{1994_Buttiker,1998_Brouwer,2004_Avron,
U1ChargePump,2010_Braunlich}
\be
\label{Eq:1Winding}
\nu_1(R)=\frac{1}{2\pi i}\int_{X=S^1} \text{Tr}\left(R^\dag \dd R
\right) \in \mathbb Z.
\ee
In this work, we propose that higher Berry invariants can be detected in a completely analogous manner. For a gapped system defined over a three-dimensional parameter space $X$, taken to be a closed manifold, we introduce the higher winding number, expressed in terms of the boundary reflection matrix, as
\begin{equation}
\label{Eq:3Winding}
\boxed{
    \nu_3(R) = \frac{1}{24 \pi^2} \int_{X} \text{Tr} (R^\dagger \dd R)^{\wedge 3}\in \mathbb Z
    }\,.
\end{equation}

This expression is motivated by two key observations.

\smallskip

First, given a smooth map as defined in \eqref{eq:map}, with 
$X$ a $d$-dimensional closed manifold, one can always define a $d$-dimensional winding number \cite{2023spectral}. The integer in Eq.~\eqref{Eq:3Winding} is precisely the homotopy class of the map in \eqref{eq:map}.
\footnote{See also the recent work \cite{2024_Ken_discreteWinding}, which studies the quantization of $\nu_3$, even when $X$ is replaced by a discrete approximation.}

Second, previous works \cite{prodan2016bulk,
2011_Akhmerov,
Beenakker_2011,Wimmer_2011,
Fulga_2012,2017_Ryu,Schulz_Baldes_2020,2011_Gurarie} have demonstrated that, for free-fermion gapped Hamiltonians in general dimensions with fixed parameters, 
the complete bulk topological information is encoded in the boundary/surface scattering matrix. For instance, if one imposes translation invariance along the spatial directions parallel to the surface and interprets the parameter 
$\lambda$ as the crystal momentum $k$ in those directions, then \eqref{Eq:3Winding} reduces to the boundary invariant associated with the second Chern number of four-dimensional topological insulators. 
In the present work, however, we treat $\lambda$ as an external control parameter. The bulk-boundary correspondence (for boundary scattering) established in \cite{prodan2016bulk,
2011_Akhmerov,
Beenakker_2011,Wimmer_2011,
Fulga_2012,2017_Ryu,Schulz_Baldes_2020,2011_Gurarie} can be naturally extended to such parametrized families of gapped free-fermion systems \cite{Schulz_Baldes_2020}. This extension leads to our boundary-scattering characterization of higher Berry phases in one-dimensional systems.

We study this boundary-scattering approach in both continuum field theories and lattice models. In particular, for lattice realizations we demonstrate that the topological invariant defined in \eqref{Eq:3Winding} remains robust against perturbations such as disorder.

The remainder of this paper is organized as follows.
In Sec.~\ref{Sec:FieldTheory}, we begin with a field-theoretic analysis of the setup. We then turn to explicit lattice realizations in Sec.~\ref{Sec:Lattice}, where we investigate both exactly solvable models and disordered models that are not analytically tractable.
In Sec.~\ref{Sec:Discuss}, we conclude with a discussion of possible generalizations to parametrized gapped systems in higher spatial dimensions. Several appendices provide technical details, including explicit calculations of the reflection matrices for the lattice models.

\section{Field theory analysis}
\label{Sec:FieldTheory}

Before analyzing lattice realizations, it is useful to first understand how the formulas in \eqref{Eq:1Winding} and \eqref{Eq:3Winding} are implemented in field theory. To this end, we revisit the examples recently discussed in \cite{Hsin_2020,wen_2025}.

\subsection{Detect $U(1)$ Thouless pump}

We consider the setup shown in Fig.~\ref{Fig:scheme}, where the gapped system is described by a massive Dirac fermion with a global $U(1)$ symmetry, and the parameter space is $X=S^1$. The parametrized Hamiltonian takes the form
\begin{equation}
\label{H_U1}
H_{\text{gap}}=\int_0^{+\infty}\dd x\,\Psi^\dag(x)
\big(-i\sigma_3 \partial_x+m_0 \sigma_1+m_1\sigma_2\big)\Psi(x),
\end{equation}
where $\Psi(x)=(\psi_R(x),\,\psi_L(x))^T$ is the Dirac fermion operator with right- and left-moving components, and $\sigma_i$ are Pauli matrices acting in the chiral basis.
The mass parameters $(m_0,m_1)$ are chosen to lie on the circle $X=S^1$, which we parametrize as
\begin{equation}
\label{X_S1}
m_1+im_0 = m\, e^{i\alpha},
\qquad \alpha\in[0,2\pi],
\end{equation}
with fixed amplitude $m>0$. As $\alpha$ is adiabatically varied from $0$ to $2\pi$, the system realizes a quantized charge pump along the one-dimensional bulk, i.e., the Thouless pump \cite{thouless_1983}.
For the lead in Fig.~\ref{Fig:scheme}, we take a gapless Dirac fermion of the same form as in \eqref{H_U1}, but with the mass term set to zero. Following the method in \cite{wen_2025}, one finds that the boundary reflection condition at $x=0$ is
\begin{equation}
\label{U1_boundary}
\psi_{\text{out}}(x=0)=e^{-i\alpha}\,\psi_{\text{in}}(x=0),
\qquad \alpha\in[0,2\pi].
\end{equation}
Thus, the reflection matrix is simply
\begin{equation}
R=e^{-i\alpha}.
\end{equation}
As $\alpha$ winds once around $S^1$, the reflection phase acquires a nontrivial winding, yielding the quantized invariant $\nu_1=-1$ according to \eqref{Eq:1Winding}. This boundary winding number faithfully captures the bulk topological response, i.e., the quantized charge pumping in the gapped system.

\subsection{Detect higher Berry invariant}

Now let us consider a gapped system with a nontrivial higher Berry phase. The parametrized gapped Hamiltonians we will consider in Fig.\ref{Fig:scheme} are \cite{Hsin_2020,wen_2025}
\begin{equation}
\label{H_S3}
\small
    H_{\text{gap}} = \int_0^{+\infty}
    \dd x\,\big[-i (\mathbb I\otimes\sigma_3) \partial_x + m_0 (\mathbb I\otimes\sigma_1) + \sum_{j=1}^3 m_j (\sigma_j \otimes \sigma_2)\big],
\end{equation}
where the parameter space is taken to be $X=S^3$, subject to the constraint $\sum_{\mu=0}^3 m_\mu^2 = 1$.

Physically, as the parameters are varied over $X=S^3$, the system exhibits a flow of the conventional two-form Berry curvature, resulting in a quantized Chern-number pump along the one-dimensional system \cite{2023_Wen,2024_Sommer1,wen_2025}.

For the lead in Fig.~\ref{Fig:scheme}, we take the same Hamiltonian \eqref{H_S3} but set the mass terms to zero. Solving the corresponding scattering problem yields the reflection matrix \cite{wen_2025}
\footnote{Note that, from the perspective of boundary conformal field theory (BCFT), the parametrized reflection matrices can be interpreted as twisted boundary conditions. Adiabatically changing this twisted boundary condition will induce a multi-parameter spectral flow in the BCFT \cite{wen_2025}. 
In the present work, however, we extract the topological information directly from the reflection matrices, without analyzing the spectral flow.
}
\begin{equation}
R = m_0 \mathbb I + i \sum_{k=1}^3 m_k \sigma_k
\in SU(2).
\end{equation}
To parametrize $S^3$, we introduce
\begin{equation}\label{Eq:ParametrizeS3}
\begin{aligned}
m_0 &= \cos\alpha,\quad
m_1 = \sin\alpha\,\cos\theta,\\
m_2 &= \sin\alpha\,\sin\theta\,\cos\phi,\quad
m_3 = \sin\alpha\,\sin\theta\,\sin\phi,
\end{aligned}
\end{equation}
with $\alpha,\theta\in[0,\pi]$ and $\phi\in[0,2\pi]$.
Substituting $R$ with this parametrization into \eqref{Eq:3Winding}, 
we find that the higher Berry invariant associated with the reflection matrix is $\nu_3[R] = -1$.
This integer quantifies the Chern-number pump along the one-dimensional gapped system
\cite{2023_Wen,2024_Sommer1,wen_2025}.

In the next section, we construct an explicit lattice realization of this field theory and compute the same topological invariant from the lattice scattering problem.

\section{Detect higher Berry phase via boundary scattering:
in lattice models}
\label{Sec:Lattice}

In this section, we implement the boundary-scattering approach to detect higher Berry phases in lattice models. We analyze both exactly solvable models and models that are not analytically tractable. While our primary focus is on systems exhibiting higher Berry phases, we also include, in Appendix~\ref{Appendix:ThoulessPump}, a complementary discussion on detecting the conventional $U(1)$ Thouless charge pump using the same approach.

\subsection{Setup and Method}

Let us first give an overview of our setup and method.
More details can be found in Appendix \ref{Appendix:Method}.

We consider a family of one-dimensional, gapped, short-range-entangled free-fermion systems described by the Hamiltonian
\begin{equation}
\label{Eq:Hamiltonian}
\begin{split}
    H_\text{gap} =& \sum_{i=0}^{2L-1} (-1)^i\psi^\dagger_{i} \vec{m}\cdot\vec{\sigma} \psi_{i} + \sum_{i=0}^{L-1}  f_+(m_0) (\psi^\dagger_{2i+1}\psi_{2i} + \text{h.c.})\\
    &+ \sum_{i=1}^L f_-(m_0)(\psi^\dagger_{2i-1}\psi_{2i} + h.c.),
    \end{split}
\end{equation}
where $\psi_i = (\psi_{i,\uparrow}, \psi_{i,\downarrow})$ contain two complex fermions, $\vec{m} = (m_1,m_2,m_3)^T$ are the mass parameters that are real, and $\vec{\sigma} = (\sigma_1,\sigma_2,\sigma_3)^T$ are the Pauli matrices.
The functions $f_+(m_0)$ and $f_-(m_0)$ represent tunable hopping amplitudes that depend on an additional parameter $m_0$. The four parameters $(m_0, \vec{m})$ define our parameter space $X = S^3$, subject to the constraint
\be
\label{Eq:S3_mass}
|\vec{m}|^2+ m_0^2 = 1.
\ee

Note that the Hamiltonian in \eqref{Eq:Hamiltonian} is fully gapped when defined on an infinite chain. However, in the presence of an open boundary, gapless boundary modes may emerge for certain regions of the parameter space.
For the lead shown in Fig.~\ref{Fig:scheme}, we consider a gapless free-fermion chain that is coupled to the gapped system. The total Hamiltonian then takes the form
\begin{equation}
H = \sum_{i<0} \frac{v_0}{2}\bigl(\psi_i^\dagger \psi_{i+1} + \text{h.c.}\bigr) + H_\text{gap},
\end{equation}
where $\psi_i = (\psi_{i,\uparrow}, \psi_{i,\downarrow})$, 
and $v_0$ denotes the group velocity in the lead. We choose the coupling between the lead and the sample to be identical to the hopping amplitude within the lead itself.
Throughout this work, we choose the Fermi energy at zero energy, which lies inside the bulk gap of $H_\text{gap}$.

For $j\leq 0$ in the lead, we assume the scattering ansatz
\begin{equation}
    \psi_j = (e^{ikj} \mathbb I + e^{-ikj} R)A_0,
\end{equation}
where $R$ is the reflection matrix and $A_0$ is an arbitrary vector that describe the spin components of the incoming wave.
This allows us to package the semi-infinite lead into a self energy $\Sigma_0 = \frac{1}{2}v_0 e^{ik}$ as well as an effective source $q_0 = -i v_0 \sin k \, A_0$ both localized at site $0$ (See Appendix \ref{Appendix:Method}).
Therefore, the wave function $\psi_j$ for this scattering problem can be solved by
\begin{equation}
    \psi_0 = G_{00}(\omega) q_0,
\end{equation}
where $G(\omega) = (\omega I - H_\text{gap}-\Sigma_0)^{-1}$ is the Green's function for the effective Hamiltonian with self energy and effective source, which is obtained by the continued fraction method on lattice.
By wave function matching, we can then solve for the reflection matrix
\begin{equation}\label{Eq:Reflection}
    R(\omega) = -\mathbb I - iv_0  \sin k \, G_{00}(\omega).
\end{equation}

\begin{figure}
\centering
\includegraphics[width=2.7in]{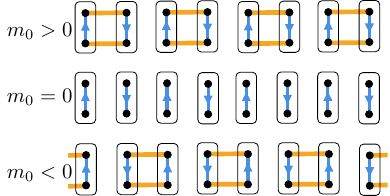}
\caption{
Exactly solvable free-fermion lattice models 
that exhibit a nontrivial higher Berry phase with parameter space $X=S^3$. Each super-site (black box) contains two sites, and 
the orange lines correspond to the hopping characterized by $m_0$. Each blue line corresponds to a hopping characterized by $\vec m=(m_1,m_2,m_3)^T$.
These parameters satisfy the constraint in \eqref{Eq:S3_mass}.
}
\label{Fig:ToyModel}
\end{figure}

\subsection{Exactly solvable model}
\label{Sec:ToyModel}

We first consider an exactly solvable model inspired by the suspension construction proposed in \cite{2023_Wen}.

As illustrated in Fig.~\ref{Fig:ToyModel}, we take the Hamiltonian in Eq.~(\ref{Eq:Hamiltonian}) and specify the functions
\begin{equation}
\label{Eq:f+}
\begin{aligned}
    f_+ = 
    \begin{cases}
        m_0,  &0\le m_0\le 1,\\
        0,  &\text{otherwise},
    \end{cases}
\end{aligned}  
\end{equation}
and
\begin{equation}
\label{Eq:f-}
\begin{aligned}
    f_- = 
    \begin{cases}
        -m_0,  &-1\le m_0\le 0,\\
        0,  &\text{otherwise}.
    \end{cases}
\end{aligned}  
\end{equation}

The physical idea underlying this construction is the following.
When $m_0=0$, the one-dimensional system reduces to a collection of decoupled zero-dimensional subsystems. The Hamiltonian for the $i$-th super-site (see Fig.\ref{Fig:ToyModel}) takes the form
\begin{equation}
H^i_{0d}= (-1)^i \psi^\dagger_{i} \, \vec{m}\cdot\vec{\sigma} \, \psi_{i}.
\end{equation}
For each super-site, the ground state carries a Chern number $\pm 1$ over the parameter space $S^2$ defined by $|\vec m|^2=1$, with the sign alternating between neighboring super-sites due to the factor $(-1)^i$.
Turning on $m_0$ couples adjacent super-sites in a dimerized pattern. For $m_0<0$, the coupling occurs between the $(2i-1)$-th and $2i$-th super-sites. In contrast, for $m_0>0$, the coupling is shifted to connect the $2i$-th and $(2i+1)$-th super-sites. This switching mechanism is essential for realizing the suspended topological structure of the model and leads to a quantized Chern-number pump, as analyzed in detail in \cite{2023_Wen}.

Due to the dimerized structure described above, the analysis of the reflection matrix reduces to a minimal two-super-site problem. It is therefore sufficient to consider the following effective Hamiltonian for the gapped system.
For $m_0> 0$, we take
\be
H_{\text{dimer}}=\psi^\dagger_0 \vec{m}\cdot\vec{\sigma} \psi_0 - \psi^\dagger_1 \vec{m}\cdot\vec{\sigma} \psi_1 + m_0 (\psi^\dagger_{1}\psi_{0} + \psi^\dagger_{0}\psi_{1} ),
\ee
while for $m_0\le 0$, the system reduces to a single decoupled super-site,
\be
H_{\text{dimer}}=\psi^\dagger_0  \, (\vec{m}\cdot\vec{\sigma} ) \, \psi_0.
\ee
In this simplified setting, the local Green’s function $G_{00}$ can be computed analytically. One finds
\begin{equation}
    G_{00} = \left\{
    \begin{split}
       & \left(-\vec{m}\cdot\vec{\sigma} - \frac{i v_0}{2}+ m_0^2 (\vec{m}\cdot\vec{\sigma})^{-1} \right)^{-1},  \quad m_0> 0\\
       & \left(-\vec{m}\cdot\vec{\sigma} - \frac{i v_0}{2} \right)^{-1}, \quad m_0\le 0\\
        \end{split}
    \right.
\end{equation}

Substituting these expressions into \eqref{Eq:Reflection}, we obtain the corresponding reflection matrix $R$. For the parametrization given in \eqref{Eq:ParametrizeS3}, the three-dimensional winding number reads
\begin{equation}
    \nu_3(R)= \frac{1}{4\pi^2} \int_X [\text{Tr}(R^\dagger \dd R)^{\wedge 3}]_{\alpha \theta \phi} \,\dd\alpha \wedge \dd\theta \wedge \dd\phi,
\end{equation}
where
\begin{equation}
    \text{Tr}(R^\dagger d R)^{\wedge 3} = 
    \left\{
    \begin{split}
&\frac{-1024\cos{\alpha}\sin^2{\alpha}\sin{\theta}}{(-9+\cos{2\alpha})^3},  \quad m_0> 0,\\
&\frac{-128\cos{\alpha}\sin^2{\alpha}\sin{\theta}}{(-3+2\cos{2\alpha})^3}, \quad m_0\le 0.
    \end{split}
    \right.
\end{equation}
The integral can then be carried out analytically, leading to the quantized higher Berry invariant
\begin{equation}
\label{nu3_qunantize}
\nu_3(R) = -1.
\end{equation}
This invariant characterizes the quantized Chern number pump
in the one dimensional gapped system \cite{2023_Wen,2024_Sommer1,wen_2025}.

\subsection{General free fermion chain}
\label{Sec:GeneralFermion}

We now turn to a more general situation by introducing a uniform hopping term, so that the system can no longer be viewed as a collection of decoupled dimers. Concretely, we modify the functions in Eqs.~\eqref{Eq:f+} and \eqref{Eq:f-} according to
\begin{equation}
\label{Eq:f'}
     f'_\pm = \frac{v_1}{2} + f_\pm,
\end{equation}
where the parameter $v_1$ controls the strength of the additional uniform hopping.
As a result, the system becomes fully connected, and the reflection matrix must be obtained by considering scattering from the entire chain rather than from an isolated dimer.

As shown in Appendix \ref{Appendix:Method}, the Green's function at the super-site $0$ for a chain $j\geq0$ can be expressed as the Green's function at site $1$ for the chain with $j\geq 1$ as follows:
\begin{equation}
    G^{j\geq 0}_{00} = \left(- \vec{m}\cdot \vec{\sigma} - \frac{v_0}{2} e^{i k} - (\frac{v_1}{2}+m_0)^2 G^{j\geq 1}_{11} \right)^{-1}.
\end{equation}
Similarly, the Green’s function for the chain with $j\geq 1$ can be written recursively in terms of that for the chain with $j\geq 2$. Iterating this procedure systematically shortens the chain and leads to a continued-fraction representation of the Green’s function.

In the thermodynamic limit $L\to\infty$, the calculation simplifies considerably. The key observation is that the semi-infinite chain becomes self-similar: removing one unit cell (two super-sites in the present model) leaves the remaining chain invariant. For instance, one has
 $G^{j\geq 1}_{11} = G^{j\geq 3}_{33}$.
As a result, $G^{j\geq 1}{11}$ can be determined by solving the following self-consistent matrix equation:
\begin{equation}
\label{G_self}
\small
    \big(G^{j\geq 1}_{11}\big)^{-1} = \vec{m}\cdot \vec{\sigma} - (\frac{v_1}{2}- m_0)^2 \left( - \vec{m}\cdot \vec{\sigma} - (\frac{v_1}{2}+m_0)^2 G^{j\geq 1}_{11} \right)^{-1}.
\end{equation}
This self-consistent equation fully determines the Green’s function and forms the basis for computing the reflection matrix of the infinite chain.

In our lattice model, we can diagonalize $G^{j\geq 1}_{11}$ by
\begin{equation}
    U(\theta, \phi) =
\begin{pmatrix}
\cos\frac{\theta}{2} & -e^{i\phi}\sin\frac{\theta}{2} \\
e^{-i\phi}\sin\frac{\theta}{2} & \cos\frac{\theta}{2}
\end{pmatrix},
\end{equation}
such that the matrix equation in \eqref{G_self} becomes two decoupled quadratic equations.
We can write down the analytic form for the reflection matrix as 
\begin{equation}
\label{R_transform}
    R = U^\dagger\begin{pmatrix}
r & 0 \\
0 & r^*
\end{pmatrix} U,
\end{equation}
with
\begin{widetext}
\begin{equation}
\label{reflection_r}
    r = \left\{
    \begin{split}
        &-\frac{m^2 + m_0 v_1 + m \sqrt{|\vec{m}|^2 + \left(m_0 + v_1 \right)^2} -i |\vec{m}| v_0}{m^2 + m_0 v_1 + m \sqrt{|\vec{m}|^2 + \left(m_0 + v_1 \right)^2} +i |\vec{m}| v_0}, \quad m_0> 0,\\
        & -\frac{-m_0^2 + |\vec{m}| ^2 + m_0 v_1 + m \sqrt{|\vec{m}|^2 + \left(m_0 - v_1 \right)^2} - i |\vec{m}| v_0}{- m_0^2 + |\vec{m}| ^2 + m_0 v_1 + m \sqrt{|\vec{m}|^2 + \left(m_0 - v_1 \right)^2} + i |\vec{m}| v_0}, \quad m_0\le 0.
        \end{split}
    \right.
\end{equation}
\end{widetext}
Based on \eqref{Eq:3Winding}, \eqref{R_transform}, and \eqref{reflection_r}, one finds that $\nu_3(R)=-1$, in agreement with the result \eqref{nu3_qunantize} obtained in the dimerized limit, as expected.

\subsection{Current for Chern number pump}

As studied in \cite{2023_Wen,2024_Sommer1,wen_2025}, the higher Berry curvature in one dimensional gapped systems characterizes the flow of ordinary Berry curvature as well as the Chern number pump.  In this section, to make the Chern-number pump more transparent, we introduce the notion of a current associated with Chern-number transport.

\medskip
We begin by illustrating the physical picture using the lattice models discussed above. 
A similar physical picture was used to study 
the Berry curvature flow in boundary conformal
field theories \cite{wen_2025}.
For the Hamiltonian in Eq.~\eqref{Eq:Hamiltonian}, one can diagonalize it in the single-particle subspace as
\begin{equation}
\label{H1_U}
    H_\text{gap} = U^\dagger(\theta, \phi) \begin{pmatrix}
        H_+(\alpha) &0\\
        0 & H_-(\alpha)
    \end{pmatrix} U(\theta, \phi),
\end{equation}
where $(\alpha,\theta,\phi)$ parametrize the three-dimensional parameter space $X=S^3$ as defined in Eq.~\eqref{Eq:ParametrizeS3}.

\begin{figure}
\centering
\includegraphics[height=4.8cm]{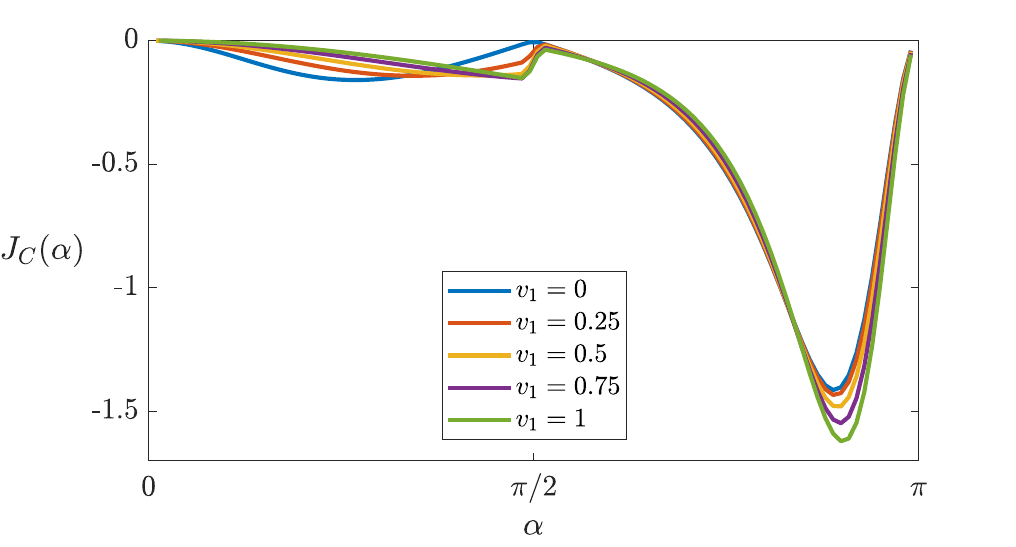}
\caption{Current of Chern number pump as a function of the pumping 
parameter $\alpha$ for different velocity $v_1$ in the gapped system
and $v_0=1$ in the lead. Here the system size for the gapped system is 
chosen as $L=20$.}
\label{Fig:compare_v1}
\end{figure}

If our one-dimensional system possesses a nonzero higher Berry invariant, then each block $H_\pm(\alpha)$ exhibits a $U(1)$ Thouless pump as the parameter $\alpha$ is varied. Notably, $H_+(\alpha)$ and $H_-(\alpha)$ pump $U(1)$ charge in opposite directions, 
so that the net $U(1)$ charge transport vanishes.
However, due to the structure in \eqref{H1_U}, each mode in $H_+(\alpha)$ ($H_-(\alpha)$) carries a quantized Chern number $+1$ ($-1$) over the two-dimensional parameter subspace $S^2_{\theta\phi}$ spanned by $\theta$ and $\phi$. Thus, although the $U(1)$ charge pumps cancel, there remains a nontrivial net flow of Berry curvature or Chern number. To characterize this flow, we define 
the 1-form Chern-number pumping current as
\begin{equation}
    J_C(\alpha) = \int_{S^2_{\theta \phi}} \text{Tr}(R^\dagger dR)^{\wedge 3}.
\end{equation}
It is noted that the integral $\int J_C(\alpha)$ is nothing but the higher winding number $\nu_3$ in \eqref{Eq:3Winding}.

As shown in Fig.~\ref{Fig:compare_v1}, we plot the distribution of $J_C(\alpha)$ for $\alpha\in[0,\pi]$ in the lattice models studied in Sec.~\ref{Sec:GeneralFermion}. For different values of the hopping strength $v_1$ in \eqref{Eq:f'}, the detailed profile of $J_C(\alpha)$ changes accordingly. However, its integral, which corresponds to the higher winding number $\nu_3$, remains invariant, taking the quantized value $-1$ in all cases considered.

\smallskip

As a remark, the detailed profile of $J_C$ in Fig.~\ref{Fig:compare_v1} also depends on the specific boundary setup. Here the lead is coupled to site $i=0$ of the gapped system, and one observes that the peak amplitude of $J_C$ appears in the region $\alpha\in[\pi/2,\pi]$. If instead the lead is attached to site $i=1$, the peak shifts to the region $\alpha\in[0,\pi/2]$. This shift reflects the underlying dimerized structure of the gapped system (See, e.g., Fig.~\ref{Fig:ToyModel}).
A similar physical picture arises in the study of 3-form higher Berry curvature in lattice systems \cite{2023_Wen,2024_Sommer1}. Depending on the location at which the Berry-curvature flow is probed, the expression for the higher Berry curvature can vary, even though the associated topological invariant remains unchanged.

\subsection{Robustness against disorder}

Now, to further examine the robustness of our topological invariant, we introduce random on-site disorder by adding the following term to the Hamiltonian defined in \eqref{Eq:Hamiltonian} and \eqref{Eq:f'}:
\begin{equation}
H_{\text{disorder}} = \sum_{i=0}^{2L-1} d_i \psi^\dagger_i \psi_i,
\end{equation}
where $d_i \in \mathbb{R}$ denotes the random on-site potential at site $i$. The disorder potentials ${d_i}$ are independently drawn from a uniform distribution on $[0, 2\bar{d}]$, with $\bar{d}$ setting the average disorder strength.
As we vary the parameters over the space $X$, the disorder configuration is kept fixed. This ensures that we are considering a continuous family of Hamiltonians throughout the parameter sweep.

\begin{figure}
\centering
\includegraphics[height=4.8cm]{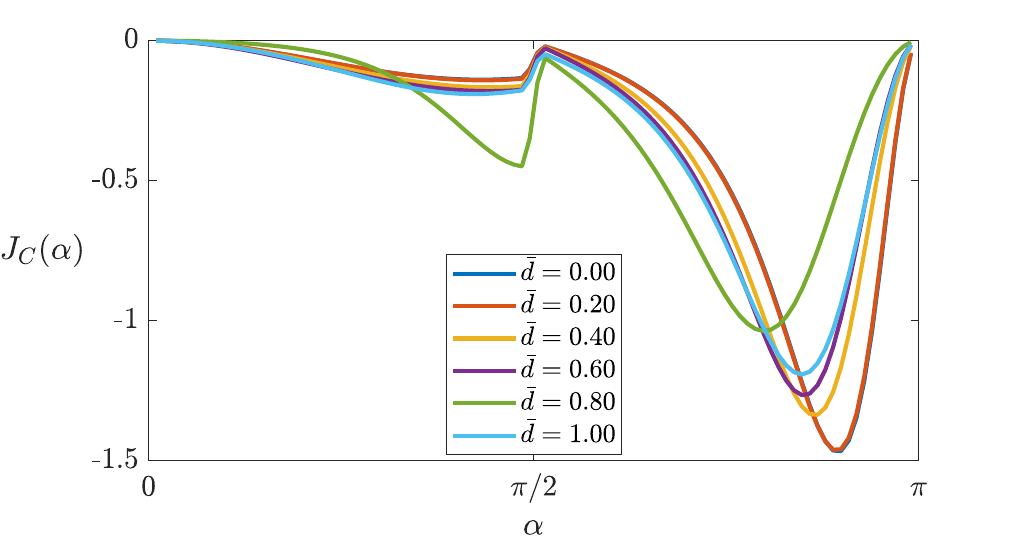}
\caption{Current of Chern number pump as a function of the pumping parameter $\alpha$ for different onsite-disorder strengths $\bar{d}$. We choose $v_0=1$, $v_1 = 0.5$, and the gapped-system size is $L=20$.}
\label{Fig:compare_d}
\end{figure}

\smallskip
As shown in Fig.~\ref{Fig:compare_d}, for different values of the disorder strength $\bar d$, the detailed profile of the Chern-number current $J_C$ changes accordingly. Nevertheless, its integral, which corresponds to the topological invariant $\nu_3$, remains quantized at $-1$, with deviations within $1\%$ in our numerical calculations\footnote{We note that if the disorder strength becomes sufficiently large, the system may undergo a phase transition. In the present analysis, we restrict the disorder strength to a regime where no phase transition occurs within the family of Hamiltonians under consideration.}. This demonstrates that the higher winding number defined in Eq.~\eqref{Eq:3Winding} is a robust topological invariant.

\section{Discussion and conclusion}
\label{Sec:Discuss}

In this work, we have proposed a boundary scattering approach to detect higher Berry invariants in one dimensional gapped free-fermion systems. We find that the topological information of the gapped system is encoded in the higher winding number of the reflection matrix, which remains robust against perturbations such as disorder.

It is natural to ask whether our formula can be extended to the case where the bulk gapped system is interacting, while the lead remains a free-fermion system \footnote{If the lead is also interacting, the situation becomes more subtle. In that case, single-particle scattering states may not be well defined, and the notion of a reflection matrix must be replaced by a boundary response formulated in terms of operators of the gapless theory. Different operator sectors can in general experience different boundary phases, so the topological information encoded in the interface may depend on which excitations are used as probes. 
}. We expect that \eqref{Eq:3Winding} continues to hold in this setting, since in one dimension the higher Berry invariant is a well-defined quantized invariant for all invertible gapped phases \cite{Kapustin2305}. In particular, when interactions are introduced into our gapped free-fermion model, the invariant should remain unchanged as long as the spectral gap stays open throughout the parameter space of the family of systems.
It would be interesting to investigate this generalization systematically, for example by adapting the approach in Ref.\cite{2003pump}.

Another interesting direction is to extend the boundary-scattering approach to higher-dimensional parametrized gapped systems. For instance, we expect that the formula in \eqref{Eq:3Winding} can be applied to the higher Thouless charge pump in $(2+1)$ dimensions \cite{KS2020_higherthouless}.
To realize a $(2+1)d$ higher Thouless pump, two external parameters are required. If the system is defined on a cylinder with translation symmetry along the direction parallel to the boundary, the corresponding momentum $k$ can be treated as a third parameter. In this way, \eqref{Eq:3Winding} may be used to detect the higher Thouless pumping invariant through boundary scattering. This perspective also suggests a close connection between the higher Thouless pump in $(2+1)d$ systems and the higher Berry phase in $(1+1)d$ systems, a relation that has indeed been explored in Ref.\cite{Kapustin2305}. It would be interesting to revisit this connection from the viewpoint of boundary physics in future work.

\acknowledgments

We thank for interesting discussions with Agnes Beaudry, Michael Hermele, Marvin Qi, Shinsei Ryu, 
Ken Shiozaki, Ophelia Sommer, and Ashvin Vishwanath.
This work is supported by a startup at Georgia 
Institute of Technology.

\appendix

\section{Details on boundary scattering}
\label{Appendix:Method}

\subsection{Reflection matrix through wave function matching}

In this appendix we summarize, following Ref.~\cite{WaveFuncMatch}, how to extract the reflection matrix from the Green's function. 
We decompose the total Hamiltonian as
\be
H = H_\text{lead} + H_\text{gap} + H_\text{int},
\ee
where $H_\text{lead}$ describes a semi-infinite lead, $H_\text{gap}$ the gapped scattering region, and $H_\text{int}$ their coupling.

We assume a uniform nearest-neighbor hopping lead. Writing the $n$-component spinor
$\psi^a_j$ where $a=1,\ldots,n$ labels the fermion species, the lead Hamiltonian is
\be
H_\text{lead}=\sum_{j=-\infty}^{-1}\frac{v_0}{2}\Bigl(\psi_{j-1}^\dagger\psi_j+\mathrm{h.c.}\Bigr).
\ee
At the interface between $j=-1$ and $j=0$, we take
\be
H_\text{int}=\frac{v_0}{2}\Bigl(\psi_{-1}^\dagger\psi_0+\mathrm{h.c.}\Bigr).
\ee
For the scattering region we consider on-site terms and nearest-neighbor hoppings,\footnote{The method extends to general quadratic Hamiltonians; however, the Green's-function construction is typically less transparent and more computationally involved.}
\be
H_\text{gap}=\sum_{j=0}^{L}\psi_j^\dagger M^j\psi_j
+\sum_{j=0}^{L-1}\Bigl(\psi_j^\dagger T^j\psi_{j+1}+\mathrm{h.c.}\Bigr),
\ee
where $M^j$ and $T^j$ are $n\times n$ matrices. For simplicity we further assume the hopping is diagonal in species,
$T^j_{ab}=t_j\delta_{ab}$.

In the single-particle sector, eigenstates in the lead ($j\le 0$) are plane waves. 
Imposing continuity at $j=0$, we write
\be\label{Eq:I+R}
\psi_j\equiv \langle j|\psi\rangle
=\Bigl(e^{ikj}\mathbb I+e^{-ikj}R\Bigr)A_0,\qquad (j\le 0),
\ee
where $R$ is the $n\times n$ reflection matrix and $A_0$ is an $n$-component unit vector specifying the incident channel.
Evaluating \eqref{Eq:I+R} at $j=0$ gives
\be\label{Eq:psi0IR}
\psi_0=(\mathbb I+R)A_0,
\ee
and at $j=-1$ yields
\be\label{Eq:psim1}
\psi_{-1}
=(e^{-ik}\mathbb I+e^{ik}R)A_0
=e^{ik}\psi_0-2i\sin k\,A_0.
\ee
The Schr\"odinger equation at $j=0$ reads
\be\label{Eq:Sch0}
0=\bra{0}(\omega\mathbb I-H)\ket{\psi}
=(\omega\mathbb I-M^0)\psi_0-\frac{v_0}{2}\psi_{-1}-t_0\psi_1,
\ee
where we used $T^0=t_0\mathbb I$. Substituting \eqref{Eq:psim1} into \eqref{Eq:Sch0} gives
\be\label{Eq:Sch0SE}
\Bigl(\omega\mathbb I-M^0-\Sigma_L\mathbb I\Bigr)\psi_0-t_0\psi_1=q_0,
\ee
with the (retarded) lead self-energy
\be
\Sigma_L=\frac{v_0}{2}e^{ik},
\ee
and the source term
\be
q_0=-i v_0\sin k\,A_0.
\ee
Therefore the problem reduces to an inhomogeneous equation on the $j\ge 0$ subspace,
\be\label{Eq:Inhom}
(\omega\mathbb I-H')\ket{\psi}=\ket{q},
\ee
where
\be
H'_{ij}=(H_\text{gap})_{ij}+\Sigma_L\,\delta_{i0}\delta_{j0},\quad
q_i=\delta_{i0}\,q_0,\quad (i,j\ge 0).
\ee
Defining the Green's function
\be
G(\omega)=(\omega\mathbb I-H')^{-1},
\ee
we have $\ket{\psi}=G(\omega)\ket{q}$ and thus
\be
\psi_0=G_{00}(\omega)\,q_0=-i v_0\sin k\,G_{00}(\omega)\,A_0,
\ee
where $G_{00}(\omega)\equiv \langle 0|G(\omega)|0\rangle$.
Combining this with \eqref{Eq:psi0IR} and using that $A_0$ is arbitrary, we obtain the reflection matrix
\be
R=-\mathbb I-i v_0\sin k\,G_{00}(\omega).
\ee

\subsection{Green's function on lattice}

In the previous subsection, we showed that the reflection matrix can be read off directly from the Green's function.
In this section we review the continued-fraction method for computing lattice Green's functions.

Working in the single-particle subspace, we partition the Hilbert space as
\[
\mathcal{H}=\mathcal{H}_A\oplus \mathcal{H}_B,\;\,
\mathcal{H}_A=\mathrm{span}\{\ket{0}\},\;\,
\mathcal{H}_B=\mathrm{span}\{\ket{i}\,|\, i\ge 1\}.
\]
With respect to this decomposition, the operator \(\omega\mathbb I - H'\) takes the block form
\begin{equation}
    \omega\mathbb I - H'=
    \begin{pmatrix}
        \omega\mathbb I - H'_{AA} & -H'_{AB}\\
        -H'_{BA} & \omega\mathbb I - H'_{BB}
    \end{pmatrix}.
\end{equation}
Let \(G(\omega)=(\omega\mathbb I-H')^{-1}\) denote the full Green's function. By the Schur-complement formula,
\begin{equation}
    G_{AA}(\omega)=\Bigl[\omega\mathbb I - H'_{AA} - H'_{AB}\,g_B(\omega)\,H'_{BA}\Bigr]^{-1},
\end{equation}
where we have introduced the Green's function of the decoupled \(B\)-subsystem,
\[
g_B(\omega):=(\omega\mathbb I - H'_{BB})^{-1}.
\]

For the 1D chain considered here, the blocks are
\begin{align}
    H'_{AA} &= (M^0+\Sigma_L \mathbb I)\otimes \ket{0}\bra{0},\\
    H'_{AB} &= t_0 \mathbb I\otimes \ket{0}\bra{1},\qquad
    H'_{BA}=H_{AB}^{\prime\dagger},\\
    H'_{BB} &= \sum_{j=1}^{L} M^j\otimes \ket{j}\bra{j}
    + \sum_{j=1}^{L-1} t_j \mathbb I\otimes \bigl(\ket{j}\bra{j+1}+\text{h.c.}\bigr).
\end{align}
Since \(\mathcal{H}_A\) is one-dimensional in real space, we obtain
\begin{equation}\label{Eq:GreensFunc}
\begin{split}
    \bra{0}G(\omega)\ket{0}
    &= \bra{0}G_{AA}(\omega)\ket{0}\\
    &= \Bigl[\omega\mathbb I - M^0 - \Sigma_L \mathbb I
    - t_0^2\,\bra{1}g_B(\omega)\ket{1}\Bigr]^{-1}.
\end{split}
\end{equation}
In other words, \(\bra{0}G(\omega)\ket{0}\) is determined by the matrix element
\(\bra{1}g_B(\omega)\ket{1}\) of the subsystem with site \(0\) removed.

This observation leads to a recursion. Define \(g^{(i)}(\omega)\) as the Green's function of the subsystem
restricted to sites \(i,i+1,\dots,L\). Then for \(1\le i \le L-1\),
\begin{equation}
    \bra{i}g^{(i)}(\omega)\ket{i}
    =\Bigl[\omega\mathbb I - M^i - t_i^2\,\bra{i+1}g^{(i+1)}(\omega)\ket{i+1}\Bigr]^{-1},
\end{equation}
with terminal condition (open boundary at \(L\))
\begin{equation}
    \bra{L}g^{(L)}(\omega)\ket{L}=(\omega\mathbb I - M^L)^{-1}.
\end{equation}
Substituting these expressions back inductively yields \(\bra{1}g_B(\omega)\ket{1}\), and hence
\(\bra{0}G(\omega)\ket{0}\) via Eq.~\eqref{Eq:GreensFunc}.

While this continued-fraction evaluation is straightforward numerically, it typically does not yield a compact
closed form. A major simplification occurs when the sample is translationally invariant (e.g., \(M^i=M\) and
\(t_i=t\) in the bulk), in which case the recursion reduces to a self-consistency equation for the surface Green's function.

\section{Detecting Thouless charge pump via boundary scattering}
\label{Appendix:ThoulessPump}

In this appendix, we apply the boundary-scattering approach to detect the topological invariant associated with the Thouless $U(1)$ charge pump, where the parameter space is $X=S^1$. The model and method we consider are closely related to those studied in Ref.\cite{U1ChargePump}.
This example serves as a useful illustration for understanding the detection of higher Berry phases discussed in the main text.

\smallskip
We model the (semi-infinite) lead as a free-fermion tight-binding chain with nearest-neighbor hopping,
\begin{equation}
H_{\text{lead}}
= \sum_{i=-\infty}^{-1} \frac{v_0}{2}\,\bigl(\psi_{i+1}^\dagger \psi_i + \text{h.c.}\bigr)\,.
\end{equation}
For an infinite chain (or away from the boundary), Fourier transforming $\psi_i=\int \frac{dk}{2\pi}e^{iki}\psi_k$ gives
\begin{equation}
H_{\text{lead}}
= \int_{-\pi}^{\pi}\frac{dk}{2\pi}\; \varepsilon(k)\,\psi_k^\dagger \psi_k,
\qquad
\varepsilon(k)=v_0\cos k\,.
\end{equation}
Thus the zero-energy (Fermi) points satisfy $\varepsilon(k_F)=0$, i.e., $k_F=\pm \frac{\pi}{2}$.
The group velocity is
\begin{equation}
v(k)=\frac{\partial \varepsilon(k)}{\partial k}=-v_0\sin k,
\end{equation}
so the Fermi velocity has magnitude $|v_F|=v_0$ at $k_F=\pm \frac{\pi}{2}$ (with the sign distinguishing left-/right-movers).

\subsection{$U(1)$ Thouless pump and boundary scattering }

A Thouless pump can be realized in a one-dimensional free-fermion system whose (generally complex) mass terms are controlled by an external parameter $\alpha:[0,T]\to S^1$.
When $\alpha(t)$ is varied adiabatically through one full cycle, an integer number of charges is transported across the system.
In a translation-invariant geometry this transport is captured by the change of bulk polarization over the cycle.
Equivalently, the total charge pumped per period is the first Chern number obtained by integrating the Berry curvature over the two-torus formed by the Brillouin zone and the pump parameter space $S^1$.

If the system is opened, the same topological invariant appears as spectral flow of edge states: during one adiabatic cycle, an edge level traverses the gap so that one electron is transferred from the right boundary to the left boundary.
The same physics is visible in a scattering setup where metallic leads are attached to one end of the pump.
In the insulating limit the lead electron is perfectly reflected, acquiring a parameter-dependent phase,
\begin{equation}
    \psi_{\mathrm{out}}^{\mathrm{lead}}(t)=R(\alpha(t))\,\psi_{\mathrm{in}}^{\mathrm{lead}}(t),
\end{equation}
with $R(\alpha(t))=e^{i\beta(t)}$.
The accumulated phase then produces a pumped charge given by Brouwer's formula~\cite{Brouwer_1998},
\begin{equation}
\begin{aligned}
    \Delta Q &=\frac{e}{2\pi}\int_{0}^{T} dt\; R^\dagger \partial_t R
    =\frac{e}{2\pi}\int_{0}^{T} dt\; \partial_t \beta(t)\\
    &=\frac{e}{2\pi}\bigl[\beta(T)-\beta(0)\bigr].
\end{aligned}
\end{equation}
While the detailed time dependence of $\alpha(t)$ and $\beta(t)$ may differ, over a closed cycle they have the same net winding.
Consequently, the pumped charge per period is quantized and equals the common winding number~\cite{Br_unlich_2009, PhysRevB.83.155429}.

\subsection{Thouless charge pump in the decoupled limit}

Before tackling the full sample Hamiltonian, it is useful to start with a simpler decoupled limit, in which the sample is described by
\begin{equation}\label{Eq:H_U1_dec}
\begin{split}
    H_{\text{dimer}} =& \sum_{i=0}^L (-1)^i m_1 a_{i}^\dagger \psi_i
    + \sum_{i=0}^L f_+(m_0)\bigl(\psi_{2i+1}^\dagger \psi_{2i} + \text{h.c.}\bigr)\\
    &+ \sum_{i=1}^L f_-(m_0)\bigl(\psi_{2i-1}^\dagger \psi_i + \text{h.c.}\bigr),
\end{split}
\end{equation}
where
\begin{equation}
  f_+(m_0)=
  \begin{cases}
        m_0,  & m_0\ge 0,\\
        0,    & \text{otherwise},
  \end{cases}
\end{equation}
and
\begin{equation}
  f_-(m_0)=
  \begin{cases}
        -m_0,   & m_0\le 0,\\
        0,      & \text{otherwise}.\\
  \end{cases}
\end{equation}
We consider the parameter space $X=S^1$ by restricting to the loop $m_0^2+m_1^2=m^2$ at fixed $m$.

Because $H_{\text{dimer}}$ is decoupled, the Green's function element $G_{00}$ depends only on the degrees of freedom directly coupled to site $0$. Thus, for the purpose of computing $G_{00}$, we may replace the full system by the effective local Hamiltonian
\begin{equation}
    H_\text{eff}=
    \begin{cases}
        m_1\bigl(\psi^\dagger_0 \psi_0 - \psi^\dagger_1 \psi_1\bigr) + m_0\bigl(\psi^\dagger_0 \psi_1 + \psi^\dagger_1 \psi_0\bigr), & m_0> 0,\\
        m_1 \psi^\dagger_0 \psi_0, & m_0\le 0.
    \end{cases}
\end{equation}

This decoupled system is effectively of finite size, so $G_{00}$ can be obtained directly. In the following we focus on fermions near the Fermi surface, $\omega=0$ and $k=\pi/2$.

For $m_0>0$, one finds
\begin{equation}
    G_{00}=\left(-m_1-\frac{i v_0}{2}+\frac{m_0^2}{m_1}\right)^{-1}
    = \frac{-2m_1}{2m^2+i v_0 m_1},
\end{equation}
where $m^2=m_0^2+m_1^2$. Using \eqref{Eq:Reflection}, the reflection amplitude is
\begin{equation}\label{Eq:m0>0}
    R=-\frac{2m^2-i v_0 m_1}{2m^2+i v_0 m_1}.
\end{equation}

For $m_0\le 0$, we instead have
\begin{equation}
    G_{00}=\left(-m_1-\frac{i v_0}{2}\right)^{-1}
    = \frac{-2}{2m_1+i v_0},
\end{equation}
and therefore
\begin{equation}\label{Eq:m0<0}
    R=-\frac{2m_1-i v_0}{2m_1+i v_0}.
\end{equation}

\begin{figure}[t]
    \centering
    \includegraphics[height=6cm]{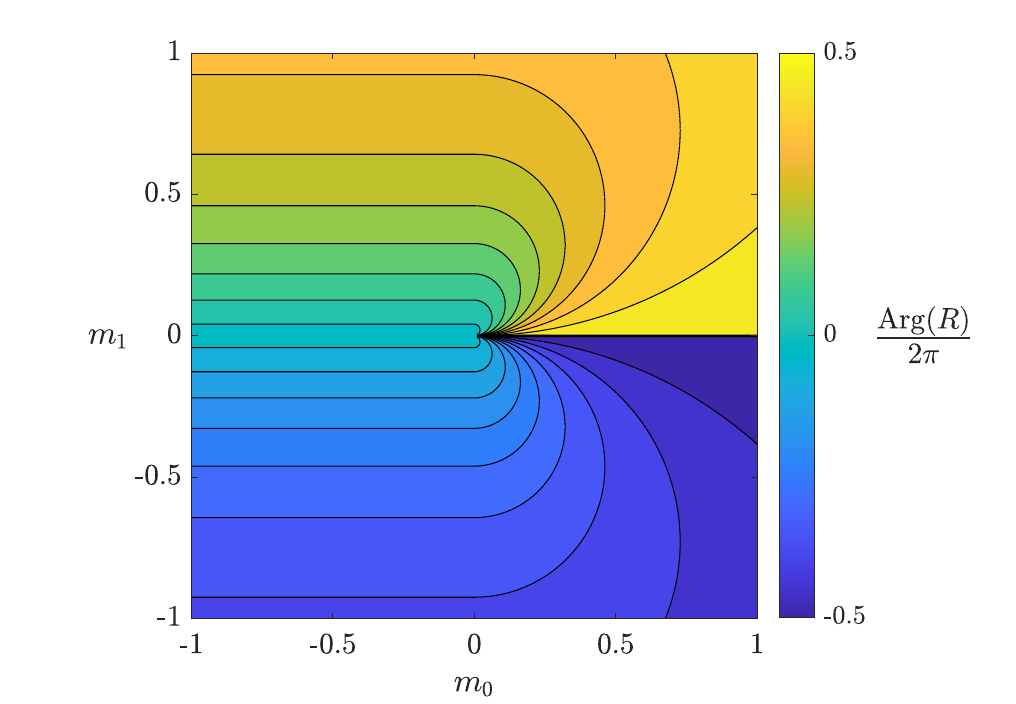}
    \caption{$\frac{1}{2\pi}\mathrm{Arg}(R)$ for different values of $m_0$ and $m_1$ in the decoupled limit, with lead velocity $v_0=1$.}
    \label{Fig:U1_dec}
\end{figure}

Since $|R|=1$, the contour integral over a loop $C$ enclosing the origin.
$\oint_C R^{-1}\, dR$ reduces to the net winding of $\mathrm{Arg}(R)$ along $C$. This integer winding equals the number of charges pumped across the system during one cycle of the parameter variation.
As shown in Fig.\ref{Fig:U1_dec}, the reflection phase winds by a full $2\pi$ for any loop that circles the origin once.

\begin{figure}[tp]
    \centering
    \includegraphics[height=5cm]{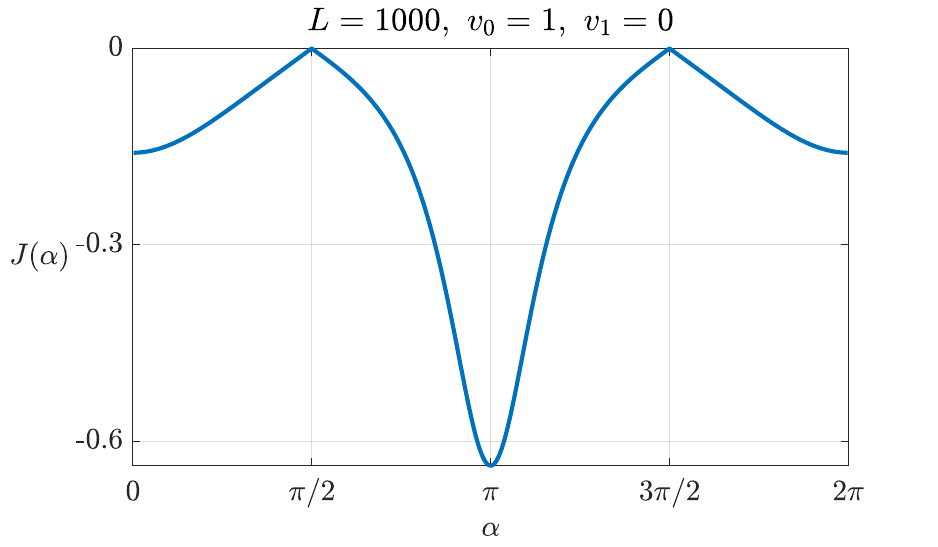}
    \caption{$J(\alpha)$ as a function of $\alpha$ in the decoupled limit, plotted according to \eqref{J_U(1)}.}
    \label{Fig:U1_dec_int}
\end{figure}

To make this explicit, we parametrize the loop as $(m_0,m_1)=(\cos\alpha,\sin\alpha)$. The reflection phase then generates a pumped current \cite{Brouwer_1998}
\begin{equation}
    J(\alpha)=\frac{1}{2\pi}R^\dagger(\alpha)\,\partial_\alpha R(\alpha),
\end{equation}
where we set the electron charge to $e=1$. This gives
\begin{equation}
\label{J_U(1)}
    J(\alpha)=
    \begin{cases}
        \dfrac{4\cos\alpha}{(-9+\cos 2\alpha)\pi}, & m_0>0,\\[6pt]
        \dfrac{2\cos\alpha}{(1+4\sin^2\alpha)\pi}, & m_0\le 0.
    \end{cases}
\end{equation}
As shown in Fig.~\ref{Fig:U1_dec_int}, we give a plot 
of $J(\alpha)$ as a function of $\alpha\in[0,\pi]$ 
according to the above equation.
Integrating over one full cycle gives an exactly quantized result,
\begin{equation}
    \Delta Q=\int_{0}^{2\pi} J(\alpha)\,d\alpha=-1,
\end{equation}
indicating that one unit of charge is pumped per loop in parameter space.

\subsection{Charge pump with hoppings}

Next, we turn on the group velocity $v_1$ within the sample by adding uniform hopping terms to Eq.~(\ref{Eq:H_U1_dec})
\begin{equation}
H_\text{gap} =  H_\text{dimer} + \sum_{i=0}^{2L-1} \frac{v_1}{2} (\psi^\dagger_{i+1}\psi_{i} + \text{h.c.}) 
\end{equation}
Again, we consider the low energy fermions with $\omega = 0$ and $k = \frac{\pi}{2}$.

For $m_0 >0$
\begin{align}
    G^{(0)}_{00} &= \left( - m_1 - \frac{i v_0}{2} - (\frac{v_1}{2} + m_0)^2 G^{(1)}_{11} \right)^{-1},\\
    G^{(1)}_{11} &= \left(  m_1 - \frac{v_1^2}{4} G^{(2)}_{22} \right)^{-1},\\
    G^{(2)}_{22} &= \left(  - m_1  - (\frac{v_1}{2} + m_0)^2 G^{(3)}_{33} \right)^{-1}.
\end{align}
Since our Hamiltonian has translational symmetry with unit cell of $2$ sites, in the $L \rightarrow \infty$ limit we have
$G^{(1)}_{11} = G^{(3)}_{33}$.
Solving for the quadratic equation, we get
\begin{equation}
    G^{(1)}_{11} = \frac{2\left(m_0^2 - m_1 ^2 + m_0 v_1  + m \sqrt{m_1^2 + \left(m_0 + v_1 \right)^2} \right)}{\left(v_1 + 2 m_0 \right)^2  m_1},
\end{equation}
where $m = \sqrt{m_0^2 + m_1^2}$, and we choose the solution that has a smooth limit at $v_1=0$.
In order to matches the result in the decoupled limit $v_1 = 0$, we choose the $+$ solution.
Therefore,
\begin{equation}
    G^{(0)}_{00} = \frac{2m_1}{ m^2 + m_0 v_1  - m \sqrt{m_1^2 + \left(m_0 + v_1 \right)^2} +i m_1 v_0 }.
\end{equation}
The reflection coefficient is given by
\begin{equation}
    R = -\frac{m^2 + m_0 v_1 + m \sqrt{m_1^2 + \left(m_0 + v_1 \right)^2} -i m_1 v_0}{m^2 + m_0 v + m \sqrt{m_1^2 + \left(m_0 + v_1 \right)^2} +i m_1 v_0}.
\end{equation}

For $m_0 <0$, we have instead
\begin{align}
    G^{(0)}_{00} &= \left( - m_1 - \frac{i v_0}{2} - \frac{v_1^2}{4} G^{(1)}_{11} \right)^{-1},\\
    G^{(1)}_{11} &= \left(  m_1 - (\frac{v_1}{2} - m_0)^2 G^{(2)}_{22} \right)^{-1},\\
    G^{(2)}_{22} &= \left(  - m_1  - \frac{v_1^2}{4} G^{(3)}_{33} \right)^{-1}.
\end{align}
Again, in the $L \rightarrow \infty$ limit we have $G^{(1)}_{11} = G^{(3)}_{33}$, which leads to
\begin{equation}
    G^{(1)}_{11} = \frac{2 \left( -m_0^2 - m_1 ^2 + m_0 v_1  + m \sqrt{m_1^2 + \left(m_0 + v_1 \right)^2} \right) }{v_1^2  m_1},
\end{equation}
where we select the branch of the solution that remains smooth in the limit $v_1\to 0$.
Therefore,
\begin{equation}
    G^{(0)}_{00} = \frac{2m_1}{ m_0^2 - m_1 ^2 + m_0 v_1  + m \sqrt{m_1^2 + \left(m_0 + v_1 \right)^2} -i m_1 v_0 }.
\end{equation}
The reflection coefficient is given by
\begin{equation}
    R = -\frac{-m_0^2 + m_1 ^2 + m_0 v_1 + m \sqrt{m_1^2 + \left(m_0 - v_1 \right)^2} - i m_1 v_0}{- m_0^2 + m_1 ^2 + m_0 v + m \sqrt{m_1^2 + \left(m_0 - v_1 \right)^2} + i m_1 v_0}.
\end{equation}

\begin{figure}[tp]
    \centering
    \includegraphics[height=5.5cm]{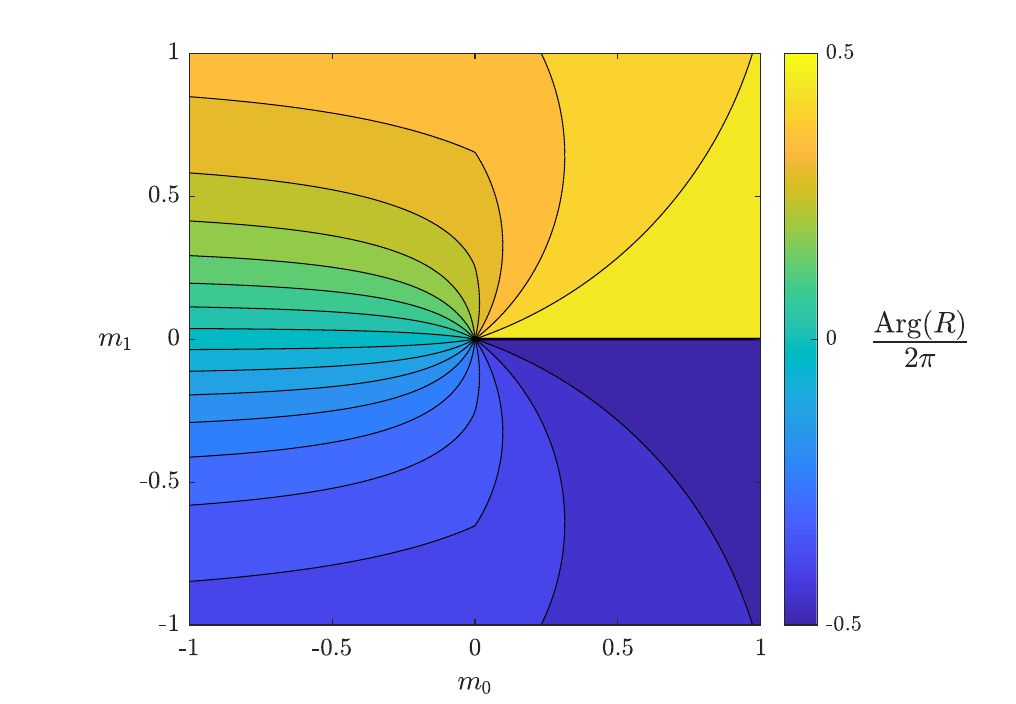}
    \caption{$\frac{1}{2\pi}\text{Arg}(R)$ for different values of $m_0$ and $m_1$ in the case where $v_0 = v_1 = 1$.}
    \label{Fig:U1_hop}
\end{figure}

As shown in Fig.~\ref{Fig:U1_hop}, the phase of reflection $\text{Arg}(R)$ will undergo a full $2\pi$ winding along 
any loop $C$ enclosing the origin, i.e., 
$
    \int_C R^{-1} dR =2\pi.
$

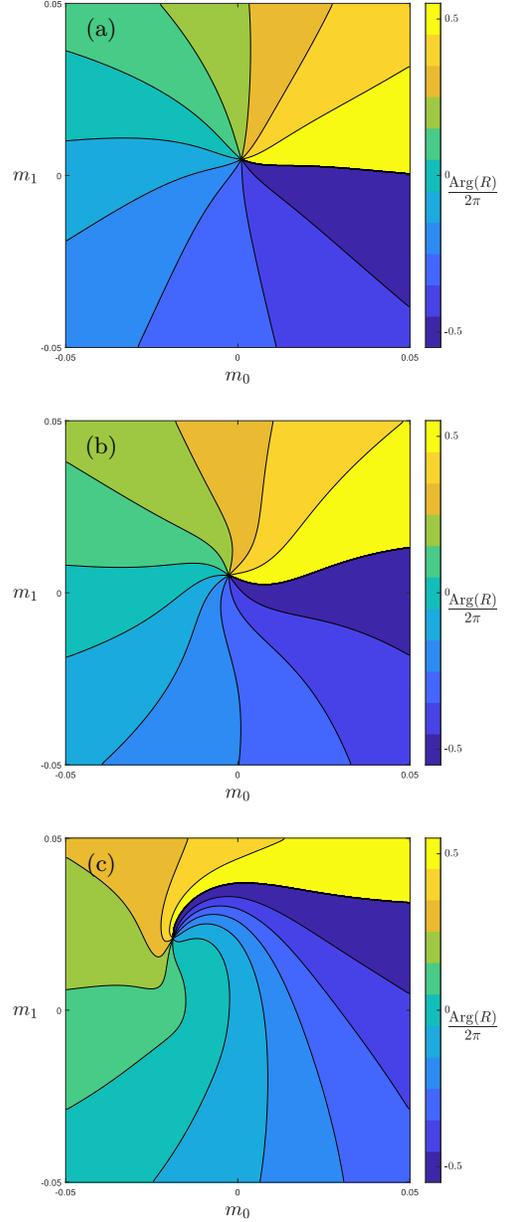
\begin{figure}[tp]
\centering
\begin{tikzpicture}
  \node[inner sep=0pt] (russell) at (0pt,0pt)
    {\includegraphics[width=3.2in]{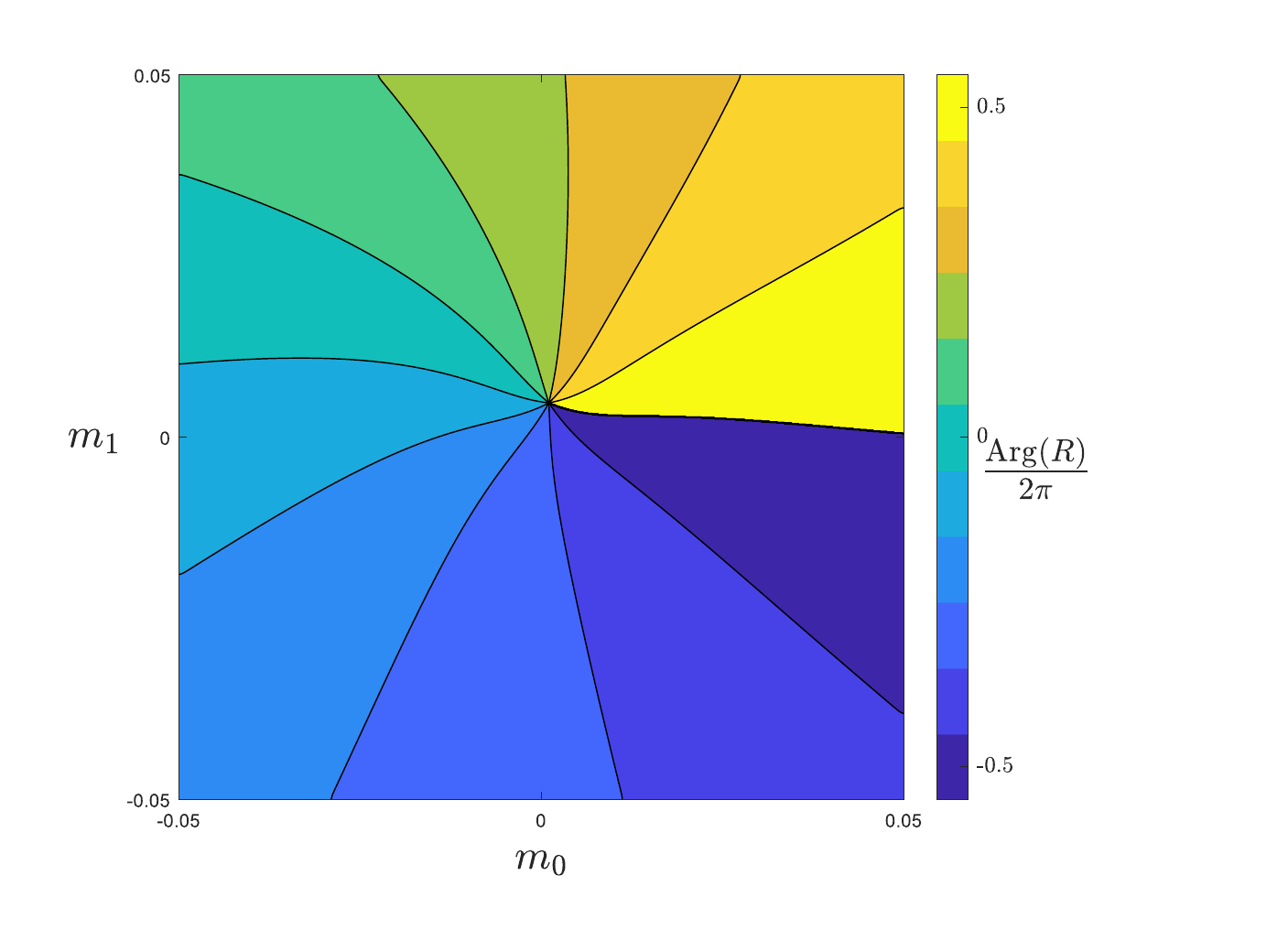}};

      \node[inner sep=0pt] (russell) at (0pt,-158pt)
    {\includegraphics[width=3.2in]{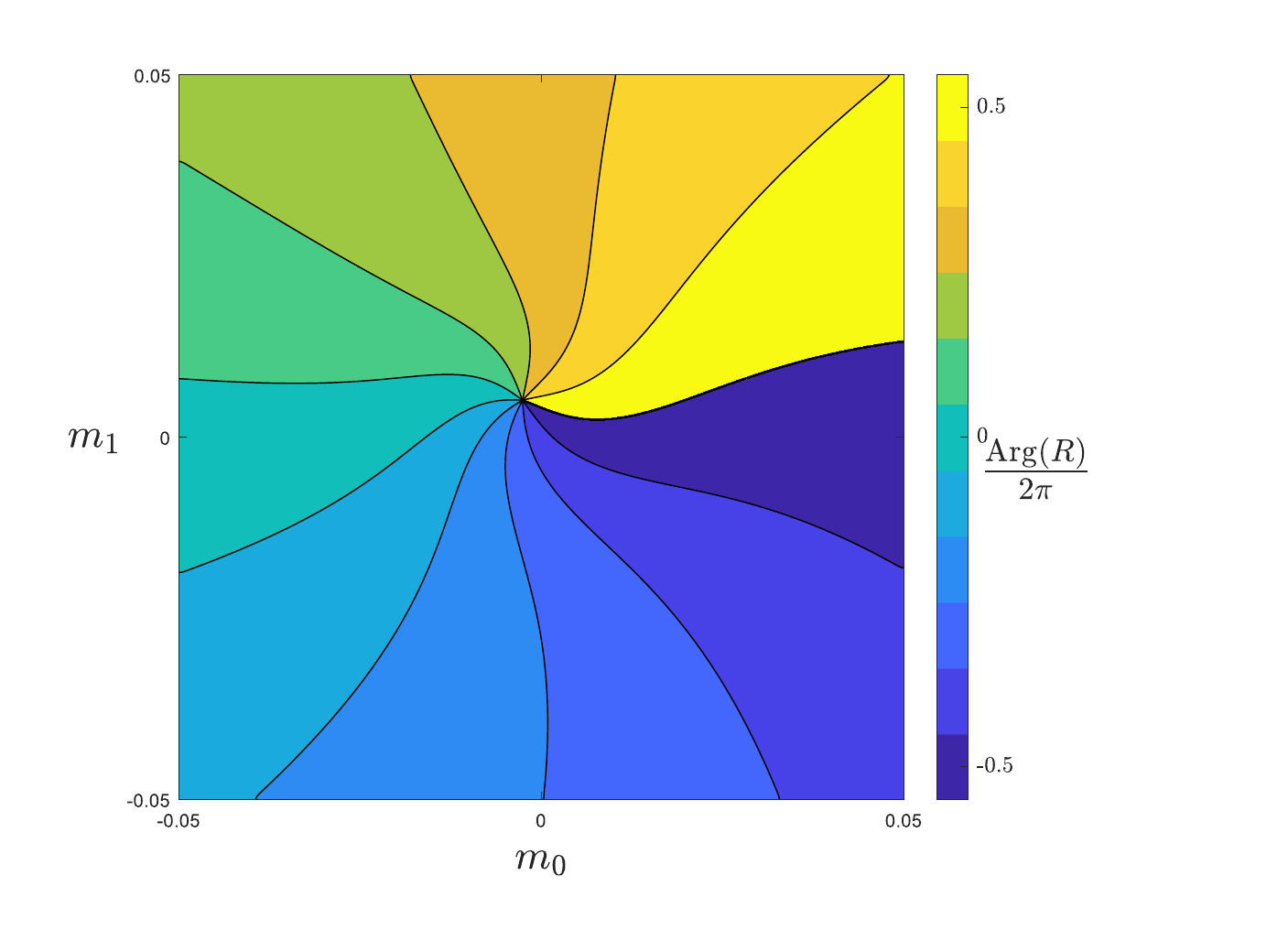}};

          \node[inner sep=0pt] (russell) at (0pt,-316pt)
    {\includegraphics[width=3.2in]{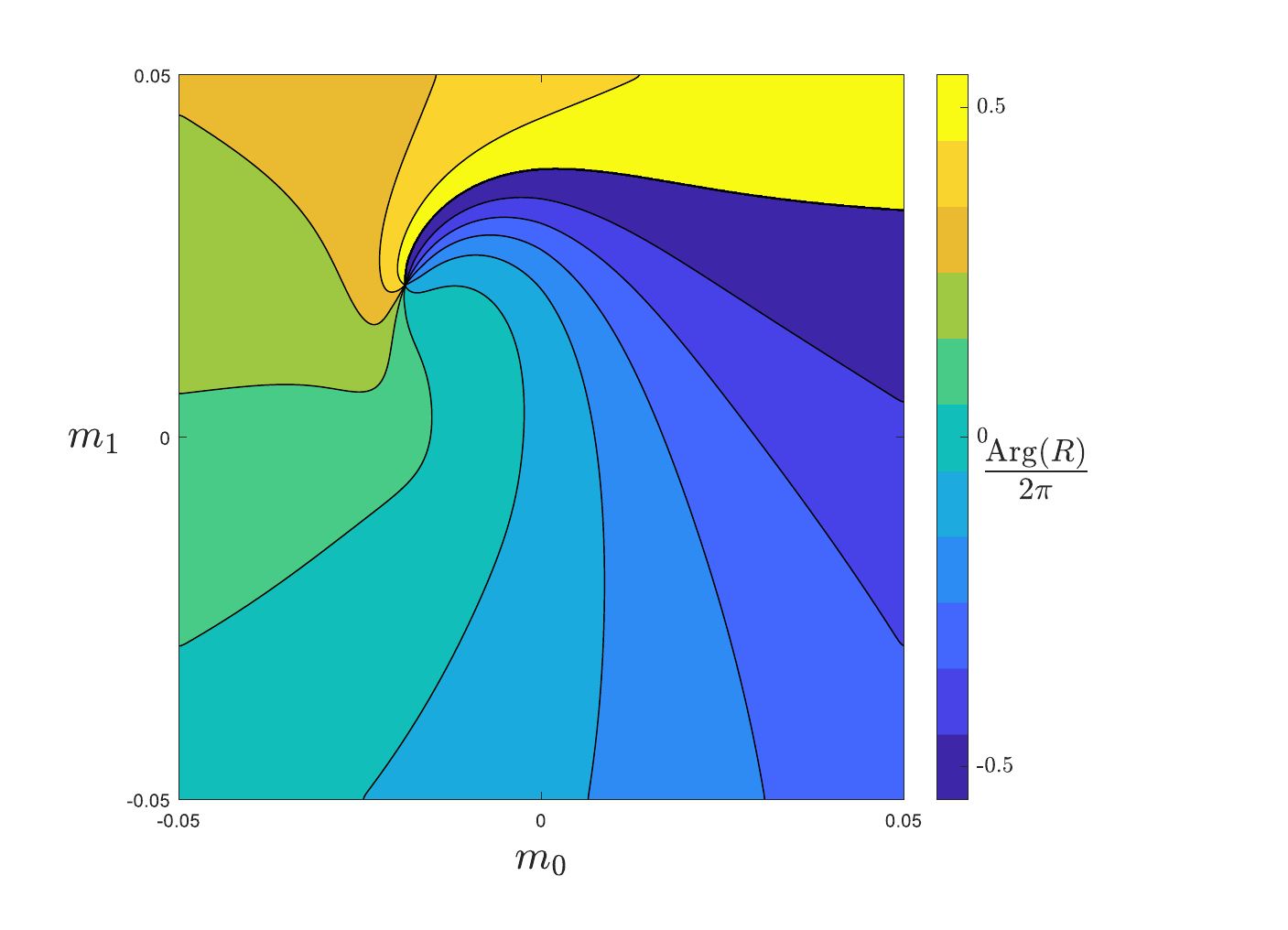}};
  
 \node at (-70pt, 60pt){\textcolor{black}{(a)}};
 \node at (-70pt, 60-158pt){\textcolor{black}{(b)}};
  \node at (-70pt, 60-316pt){\textcolor{black}{(c)}};
 
    \end{tikzpicture}
\caption{The phases $\mathrm{Arg}(R)$ for disorder strength (a) $\Bar{d} = 0.05$, (b) $\Bar{d} = 0.10$, (c) $\Bar{d} = 0.15$ with a system size $L=100,\, v_0 = 1,\, v_1 = 1$. Notice that we show the zoomed-in view near the origin to better visualize the distortion due to the disorder.}%
\label{Fig:Disorder}
\end{figure}

\subsection{Charge pump with hoppings and disorders}
In order to verify that the effect is topologically robust, we add onsite disorder to the sample Hamiltonian,
\begin{equation}
H_{\mathrm{disorder}}=\sum_{j=0}^{\infty} d_j\,\psi^\dagger_{j}\psi_{j},
\end{equation}
where ${d_j}$ are random onsite potentials with mean strength $\bar d$, drawn from a uniform distribution on $[0,2\bar d]$.

We take a system size $L=100$ and solve for the wave function numerically. As shown in Fig.~\ref{Fig:Disorder},
the presence of disorder slightly shifts the locations of the sinks and distorts the phase diagram.
Nevertheless, the topological signature remains unchanged: $\mathrm{Arg}(R)$ winds by $2\pi$ along any sufficiently large closed loop that encloses a sink, demonstrating robustness against this class of disorder.

\smallskip

As a final remark, in Figs.~\ref{Fig:U1_dec}, \ref{Fig:U1_hop}, and \ref{Fig:Disorder}, 
one can clearly observe singularities in $\text{Arg}(R)$. These singularities correspond to gapless points in the phase diagram, whose general properties have been the subject of extensive recent study.


\bibliography{Ref_Scatter}

@misc{2003pump,
      title={Adiabatic charge pumping in open quantum systems}, 
      author={J. E. Avron and A. Elgart and G. M. Graf and L. Sadun and K. Schnee},
      year={2003},
      eprint={math-ph/0209029},
      archivePrefix={arXiv},
      primaryClass={math-ph},
      url={https://arxiv.org/abs/math-ph/0209029}, 
}

@article{2010_Teo_Kane,
  title = {Topological defects and gapless modes in insulators and superconductors},
  author = {Teo, Jeffrey C. Y. and Kane, C. L.},
  journal = {Phys. Rev. B},
  volume = {82},
  issue = {11},
  pages = {115120},
  numpages = {26},
  year = {2010},
  month = {Sep},
  publisher = {American Physical Society},
  doi = {10.1103/PhysRevB.82.115120},
  url = {https://link.aps.org/doi/10.1103/PhysRevB.82.115120}
}

@article{prodan2016bulk,
  title={Bulk and boundary invariants for complex topological insulators},
  author={Prodan, Emil and Schulz-Baldes, Hermann},
  journal={K},
  year={2016},
  publisher={Springer}
}

@article{2011_Gurarie,
  title = {Bulk-boundary correspondence of topological insulators from their respective Green's functions},
  author = {Essin, Andrew M. and Gurarie, Victor},
  journal = {Phys. Rev. B},
  volume = {84},
  issue = {12},
  pages = {125132},
  numpages = {10},
  year = {2011},
  month = {Sep},
  publisher = {American Physical Society},
  doi = {10.1103/PhysRevB.84.125132},
  url = {https://link.aps.org/doi/10.1103/PhysRevB.84.125132}
}

@article{2003_Ji,
  title={An electronic mach--zehnder interferometer},
  author={Ji, Yang and Chung, Yunchul and Sprinzak, D and Heiblum, Moty and Mahalu, Diana and Shtrikman, Hadas},
  journal={Nature},
  volume={422},
  number={6930},
  pages={415--418},
  year={2003},
  publisher={Nature Publishing Group UK London}
}

@ARTICLE{2510_Yabo,
       author = {{Li}, Yabo and {Dell'acqua}, Matteo and {Mitra}, Aditi},
        title = "{Classification of Thouless pumps with non-invertible symmetries and implications for Floquet phases}",
      journal = {arXiv e-prints},
         year = 2025,
        month = oct,
          eid = {arXiv:2510.01626},
        pages = {arXiv:2510.01626},
          doi = {10.48550/arXiv.2510.01626},
archivePrefix = {arXiv},
       eprint = {2510.01626},
 primaryClass = {cond-mat.str-el},
       adsurl = {https://ui.adsabs.harvard.edu/abs/2025arXiv251001626L}
}

@ARTICLE{2024_Ken_discreteWinding,
       author = {{Shiozaki}, Ken},
        title = "{A discrete formulation for three-dimensional winding number}",
      journal = {arXiv e-prints},
         year = 2024,
        month = mar,
          eid = {arXiv:2403.05291},
        pages = {arXiv:2403.05291},
          doi = {10.48550/arXiv.2403.05291},
archivePrefix = {arXiv},
       eprint = {2403.05291},
 primaryClass = {cond-mat.mes-hall},
       adsurl = {https://ui.adsabs.harvard.edu/abs/2024arXiv240305291S}
}

@ARTICLE{2026_Inamura_a,
       author = {{Inamura}, Kansei and {Ohyama}, Shuhei},
        title = "{Generalized cluster states in 2+1d: non-invertible symmetries, interfaces, and parameterized families}",
      journal = {arXiv e-prints},
         year = 2026,
        month = jan,
          eid = {arXiv:2601.08615},
        pages = {arXiv:2601.08615},
archivePrefix = {arXiv},
       eprint = {2601.08615},
 primaryClass = {cond-mat.str-el},
       adsurl = {https://ui.adsabs.harvard.edu/abs/2026arXiv260108615I}
}

@misc{Choi_2026,
      title={Higher Connection in Open String Field Theory}, 
      author={Yichul Choi},
      year={2026},
      eprint={2602.13627},
      archivePrefix={arXiv},
      primaryClass={hep-th},
      url={https://arxiv.org/abs/2602.13627}, 
}

@misc{Brennan_2026,
      title={Generalized Families of QFTs}, 
      author={T. Daniel Brennan and Kenneth Intriligator},
      year={2026},
      eprint={2602.09105},
      archivePrefix={arXiv},
      primaryClass={hep-th},
      url={https://arxiv.org/abs/2602.09105}, 
}

@ARTICLE{2026_Else,
       author = {{Manjunath}, Naren and {Else}, Dominic V.},
        title = "{In search of diabolical critical points}",
      journal = {arXiv e-prints},
         year = 2026,
        month = jan,
          eid = {arXiv:2601.10783},
        pages = {arXiv:2601.10783},
          doi = {10.48550/arXiv.2601.10783},
archivePrefix = {arXiv},
       eprint = {2601.10783},
 primaryClass = {cond-mat.str-el},
       adsurl = {https://ui.adsabs.harvard.edu/abs/2026arXiv260110783M}
}

@article{Wimmer_2011,
   title={Quantum point contact as a probe of a topological superconductor},
   volume={13},
   ISSN={1367-2630},
   url={http://dx.doi.org/10.1088/1367-2630/13/5/053016},
   DOI={10.1088/1367-2630/13/5/053016},
   number={5},
   journal={New Journal of Physics},
   publisher={IOP Publishing},
   author={Wimmer, M and Akhmerov, A R and Dahlhaus, J P and Beenakker, C W J},
   year={2011},
   month=may, pages={053016} }

@article{Beenakker_2011,
  title = {Quantized Conductance at the Majorana Phase Transition in a Disordered Superconducting Wire},
  author = {Akhmerov, A. R. and Dahlhaus, J. P. and Hassler, F. and Wimmer, M. and Beenakker, C. W. J.},
  journal = {Phys. Rev. Lett.},
  volume = {106},
  issue = {5},
  pages = {057001},
  numpages = {4},
  year = {2011},
  month = {Jan},
  publisher = {American Physical Society},
  doi = {10.1103/PhysRevLett.106.057001},
  url = {https://link.aps.org/doi/10.1103/PhysRevLett.106.057001}
}

@ARTICLE{2026_Inamura_b,
       author = {{Ohyama}, Shuhei and {Inamura}, Kansei},
        title = "{Parameterized families of 2+1d $G$-cluster states}",
      journal = {arXiv e-prints},
         year = 2026,
        month = jan,
          eid = {arXiv:2601.08616},
        pages = {arXiv:2601.08616},
archivePrefix = {arXiv},
       eprint = {2601.08616},
 primaryClass = {cond-mat.str-el},
       adsurl = {https://ui.adsabs.harvard.edu/abs/2026arXiv260108616O}
}

@ARTICLE{Hsin_2020,
       author = {{Hsin}, Po-Shen and {Kapustin}, Anton and {Thorngren}, Ryan},
        title = "{Berry phase in quantum field theory: Diabolical points and boundary phenomena}",
      journal = {Phys. Rev. B},
         year = 2020,
        month = dec,
       volume = {102},
       number = {24},
          eid = {245113},
        pages = {245113},
          doi = {10.1103/PhysRevB.102.245113},
archivePrefix = {arXiv},
       eprint = {2004.10758},
 primaryClass = {cond-mat.str-el},
       adsurl = {https://ui.adsabs.harvard.edu/abs/2020PhRvB.102x5113H}
}

@ARTICLE{2023_Wen,
       author = {{Wen}, Xueda and {Qi}, Marvin and {Beaudry}, Agn{\`e}s and {Moreno}, Juan and {Pflaum}, Markus J. and {Spiegel}, Daniel and {Vishwanath}, Ashvin and {Hermele}, Michael},
        title = "{Flow of higher Berry curvature and bulk-boundary correspondence in parametrized quantum systems}",
      journal = {Phys. Rev. B},
         year = 2023,
        month = sep,
       volume = {108},
       number = {12},
          eid = {125147},
        pages = {125147},
          doi = {10.1103/PhysRevB.108.125147},
archivePrefix = {arXiv},
       eprint = {2112.07748},
 primaryClass = {cond-mat.str-el},
       adsurl = {https://ui.adsabs.harvard.edu/abs/2023PhRvB.108l5147W}
}

@incollection{Kapustin_2025,
title = {Topological Phases of Matter and Homotopy Theory},
editor = {Richard Szabo and Martin Bojowald},
booktitle = {Encyclopedia of Mathematical Physics (Second Edition)},
publisher = {Academic Press},
edition = {Second Edition},
address = {Oxford},
pages = {106-110},
year = {2025},
isbn = {978-0-323-95706-9},
doi = {https://doi.org/10.1016/B978-0-323-95703-8.00048-3},
url = {https://www.sciencedirect.com/science/article/pii/B9780323957038000483},
author = {Anton Kapustin}
}

@ARTICLE{2017_Ryu,
       author = {{Cho}, Gil Young and {Shiozaki}, Ken and {Ryu}, Shinsei and {Ludwig}, Andreas W.~W.},
        title = "{Relationship between symmetry protected topological phases and boundary conformal field theories via the entanglement spectrum}",
      journal = {Journal of Physics A Mathematical General},
         year = 2017,
        month = jul,
       volume = {50},
       number = {30},
          eid = {304002},
        pages = {304002},
          doi = {10.1088/1751-8121/aa7782},
archivePrefix = {arXiv},
       eprint = {1606.06402},
 primaryClass = {cond-mat.str-el},
       adsurl = {https://ui.adsabs.harvard.edu/abs/2017JPhA...50D4002C}
}

@article{WaveFuncMatch,
  title={Calculating Scattering Matrices by Wave Function Matching},
  author={G. Brocks, V. M. Karpan and P. J. Kelly, P. A. Khomyakov and I. Marushchenko and A. Starikov, M. Talanana (Twente, The Netherlands) and I. Turek (Brno, Czech Republic) and K. Xia (Beijing, China) and P. X. Xu (Twente, The Netherlands) and M. Zwierzycki (Poznan, Poland) and G. E.},
  journal={Psi-K network},
  url={https://psi-k.net/download/highlights/Highlight_80.pdf},
  year={2007},
}

@ARTICLE{1994_Buttiker,
       author = {{B{\"u}ttiker}, M. and {Thomas}, H. and {Pr{\^e}tre}, A.},
        title = "{Current partition in multiprobe conductors in the presence of slowly oscillating external potentials}",
      journal = {Zeitschrift fur Physik B Condensed Matter},
         year = 1994,
        month = mar,
       volume = {94},
       number = {1-2},
        pages = {133-137},
          doi = {10.1007/BF01307664},
       adsurl = {https://ui.adsabs.harvard.edu/abs/1994ZPhyB..94..133B}
}

@ARTICLE{2024_Prakashi,
       author = {{Prakash}, Abhishodh and {Parameswaran}, S.~A.},
        title = "{Charge pumps, boundary modes, and the necessity of unnecessary criticality}",
      journal = {Phys. Rev. B},
         year = 2025,
        month = dec,
       volume = {112},
       number = {24},
          eid = {L241117},
        pages = {L241117},
          doi = {10.1103/kztj-jcyc},
archivePrefix = {arXiv},
       eprint = {2408.15351},
 primaryClass = {cond-mat.str-el},
       adsurl = {https://ui.adsabs.harvard.edu/abs/2025PhRvB.112x1117P}
}

@ARTICLE{2023_Prakash,
       author = {{Prakash}, Abhishodh and {Fava}, Michele and {Parameswaran}, S.~A.},
        title = "{Multiversality and Unnecessary Criticality in One Dimension}",
      journal = {\prl},
         year = 2023,
        month = jun,
       volume = {130},
       number = {25},
          eid = {256401},
        pages = {256401},
          doi = {10.1103/PhysRevLett.130.256401},
archivePrefix = {arXiv},
       eprint = {2209.00037},
 primaryClass = {cond-mat.str-el},
       adsurl = {https://ui.adsabs.harvard.edu/abs/2023PhRvL.130y6401P}
}

@ARTICLE{1998_Brouwer,
       author = {{Brouwer}, P.~W.},
        title = "{Scattering approach to parametric pumping}",
      journal = {Phys. Rev. B},
         year = 1998,
        month = oct,
       volume = {58},
       number = {16},
        pages = {R10135-R10138},
          doi = {10.1103/PhysRevB.58.R10135},
archivePrefix = {arXiv},
       eprint = {cond-mat/9808347},
 primaryClass = {cond-mat.mes-hall},
       adsurl = {https://ui.adsabs.harvard.edu/abs/1998PhRvB..5810135B}
}

@ARTICLE{2010_Braunlich,
       author = {{Br{\"a}unlich}, G. and {Graf}, G.~M. and {Ortelli}, G.},
        title = "{Equivalence of Topological and Scattering Approaches to Quantum Pumping}",
      journal = {Communications in Mathematical Physics},
         year = 2010,
        month = apr,
       volume = {295},
       number = {1},
        pages = {243-259},
          doi = {10.1007/s00220-009-0983-1},
archivePrefix = {arXiv},
       eprint = {0902.4638},
 primaryClass = {math-ph},
       adsurl = {https://ui.adsabs.harvard.edu/abs/2010CMaPh.295..243B}
}

@ARTICLE{2004_Avron,
       author = {{Avron}, J.~E. and {Elgart}, A. and {Graf}, G.~M. and {Sadun}, L.},
        title = "{Transport and Dissipation in Quantum Pumps}",
      journal = {Journal of Statistical Physics},
         year = 2004,
        month = aug,
       volume = {116},
       number = {1-4},
        pages = {425-473},
          doi = {10.1023/B:JOSS.0000037245.45780.e1},
archivePrefix = {arXiv},
       eprint = {math-ph/0305049},
 primaryClass = {math-ph},
       adsurl = {https://ui.adsabs.harvard.edu/abs/2004JSP...116..425A}
}

@article{U1ChargePump,
  title = {Quantum charge pumping and electric polarization in Anderson insulators},
  author = {Chern, Chyh-Hong and Onoda, Shigeki and Murakami, Shuichi and Nagaosa, Naoto},
  journal = {Phys. Rev. B},
  volume = {76},
  issue = {3},
  pages = {035334},
  numpages = {15},
  year = {2007},
  month = {Jul},
  publisher = {American Physical Society},
  doi = {10.1103/PhysRevB.76.035334},
  url = {https://link.aps.org/doi/10.1103/PhysRevB.76.035334}
}

@misc{Copetti_2025,
      title={'t Hooft Anomalies and Defect Conformal Manifolds: Topological Signatures from Modulated Effective Actions}, 
      author={Christian Copetti},
      year={2025},
      eprint={2507.15466},
      archivePrefix={arXiv},
      primaryClass={hep-th},
      url={https://arxiv.org/abs/2507.15466}, 
}

@ARTICLE{2011_Akhmerov,
       author = {{Akhmerov}, A.~R. and {Dahlhaus}, J.~P. and {Hassler}, F. and {Wimmer}, M. and {Beenakker}, C.~W.~J.},
        title = "{Quantized Conductance at the Majorana Phase Transition in a Disordered Superconducting Wire}",
      journal = {Phys. Rev. Lett.},
         year = 2011,
        month = feb,
       volume = {106},
       number = {5},
          eid = {057001},
        pages = {057001},
          doi = {10.1103/PhysRevLett.106.057001},
archivePrefix = {arXiv},
       eprint = {1009.5542},
 primaryClass = {cond-mat.mes-hall},
       adsurl = {https://ui.adsabs.harvard.edu/abs/2011PhRvL.106e7001A}
}

@misc{2025_Jones,
      title={Charge pumps, pivot Hamiltonians and symmetry-protected topological phases}, 
      author={Nick. G. Jones and Ryan Thorngren and Ruben Verresen and Abhishodh Prakash},
      year={2025},
      eprint={2507.00995},
      archivePrefix={arXiv},
      primaryClass={cond-mat.str-el},
      url={https://arxiv.org/abs/2507.00995}, 
}

@misc{kubota2025,
      title={Stable homotopy theory of invertible gapped quantum spin systems I: Kitaev's $\Omega$-spectrum}, 
      author={Yosuke Kubota},
      year={2025},
      eprint={2503.12618},
      archivePrefix={arXiv},
      primaryClass={math-ph},
      url={https://arxiv.org/abs/2503.12618}, 
}

@misc{homotopical2023,
	archiveprefix = {arXiv},
	author = {Agnes Beaudry and Michael Hermele and Juan Moreno and Markus Pflaum and Marvin Qi and Daniel Spiegel},
	eprint = {2303.07431},
	primaryclass = {math-ph},
	title = {Homotopical Foundations of Parametrized Quantum Spin Systems},
	year = {2023}}

@article{KS2020_higherberry,
	author = {Anton Kapustin and Lev Spodyneiko},
	doi = {10.1103/physrevb.101.235130},
	journal = {Physical Review B},
	month = {jun},
	number = {23},
	publisher = {American Physical Society ({APS})},
	title = {Higher-dimensional generalizations of Berry curvature},
	url = {https://doi.org/10.1103%2Fphysrevb.101.235130},
	volume = {101},
	year = 2020}

@misc{KS2020_higherthouless,
	archiveprefix = {arXiv},
	author = {Anton Kapustin and Lev Spodyneiko},
	eprint = {2003.09519},
	primaryclass = {cond-mat.str-el},
	title = {Higher-dimensional generalizations of the Thouless charge pump},
	year = {2020}}

@article{Shiozaki_2022,
	author = {Ken Shiozaki},
	doi = {10.1103/physrevb.106.125108},
	journal = {Physical Review B},
	month = {sep},
	number = {12},
	publisher = {American Physical Society ({APS})},
	title = {Adiabatic cycles of quantum spin systems},
	url = {https://doi.org/10.1103%2Fphysrevb.106.125108},
	volume = {106},
	year = 2022}

@misc{kitaev2015talk,
	author = {Alexei Kitaev},
	note = {talk at workshop \emph{Symmetry and Topology in Quantum Matter}, Institute for Pure \& Applied Mathematics, University of California Los Angeles},
	title = {Homotopy-theoretic approach to {SPT} phases in action: $\mathbb{Z}_{16}$-classification of three-dimensional superconductors},
	year = 2015}

@article{Cordova_2020_i,
	author = {Clay Cordova and Daniel Freed and Ho Tat Lam and Nathan Seiberg},
	doi = {10.21468/scipostphys.8.1.001},
	journal = {{SciPost} Physics},
	month = {jan},
	number = {1},
	publisher = {Stichting {SciPost}},
	title = {Anomalies in the space of coupling constants and their dynamical applications I},
	url = {https://doi.org/10.21468%2Fscipostphys.8.1.001},
	volume = {8},
	year = 2020}

@article{Cordova_2020_ii,
	author = {Clay Cordova and Daniel Freed and Ho Tat Lam and Nathan Seiberg},
	doi = {10.21468/scipostphys.8.1.002},
	journal = {{SciPost} Physics},
	month = {jan},
	number = {1},
	publisher = {Stichting {SciPost}},
	title = {Anomalies in the space of coupling constants and their dynamical applications {II}},
	url = {https://doi.org/10.21468%2Fscipostphys.8.1.002},
	volume = {8},
	year = 2020}

@article{Choi_Ohmori_2022,
	author = {Yichul Choi and Kantaro Ohmori},
	doi = {10.1007/jhep09(2022)022},
	journal = {Journal of High Energy Physics},
	month = {sep},
	number = {9},
	publisher = {Springer Science and Business Media {LLC}},
	title = {Higher Berry phase of fermions and index theorem},
	url = {https://doi.org/10.1007%2Fjhep09%282022%29022},
	volume = {2022},
	year = 2022}

@article{Else_2021,
	author = {Dominic V. Else},
	doi = {10.1103/physrevb.104.115129},
	journal = {Physical Review B},
	month = {sep},
	number = {11},
	publisher = {American Physical Society ({APS})},
	title = {Topological Goldstone phases of matter},
	url = {https://doi.org/10.1103%2Fphysrevb.104.115129},
	volume = {104},
	year = 2021}

@ARTICLE{2024_Multi_WF,
       author = {{Liu}, Bowei and {Zhang}, Junjia and {Ohyama}, Shuhei and {Kusuki}, Yuya and {Ryu}, Shinsei},
        title = "{Multi wavefunction overlap and multi entropy for topological ground states in (2+1) dimensions}",
      journal = {arXiv e-prints},
         year = 2024,
        month = oct,
          eid = {arXiv:2410.08284},
        pages = {arXiv:2410.08284},
          doi = {10.48550/arXiv.2410.08284},
archivePrefix = {arXiv},
       eprint = {2410.08284},
 primaryClass = {cond-mat.str-el},
       adsurl = {https://ui.adsabs.harvard.edu/abs/2024arXiv241008284L}
}

@misc{bose2025,
      title={Symmetry constrained field theories for chiral spin liquid to spin crystal transitions}, 
      author={Anjishnu Bose and Andrew Hardy and Naren Manjunath and Arun Paramekanti},
      year={2025},
      eprint={2505.01491},
      archivePrefix={arXiv},
      primaryClass={cond-mat.str-el},
      url={https://arxiv.org/abs/2505.01491}, 
}

@ARTICLE{2024_Geiko,
       author = {{Geiko}, Roman},
        title = "{Parametrized topological phases in 1d and T-duality}",
      journal = {arXiv e-prints},
         year = 2024,
        month = dec,
          eid = {arXiv:2412.20905},
        pages = {arXiv:2412.20905},
          doi = {10.48550/arXiv.2412.20905},
archivePrefix = {arXiv},
       eprint = {2412.20905},
 primaryClass = {math-ph},
       adsurl = {https://ui.adsabs.harvard.edu/abs/2024arXiv241220905G}
}

@article{Ohyama_2022,
	author = {Ohyama, Shuhei and Shiozaki, Ken and Sato, Masatoshi},
	doi = {10.1103/PhysRevB.106.165115},
	issue = {16},
	journal = {Phys. Rev. B},
	month = {Oct},
	numpages = {37},
	pages = {165115},
	publisher = {American Physical Society},
	title = {Generalized Thouless pumps in $(1+1)$-dimensional interacting fermionic systems},
	url = {https://link.aps.org/doi/10.1103/PhysRevB.106.165115},
	volume = {106},
	year = {2022}}

@misc{ohyama2023discrete,
	archiveprefix = {arXiv},
	author = {Shuhei Ohyama and Yuji Terashima and Ken Shiozaki},
	eprint = {2303.04252},
	primaryclass = {cond-mat.str-el},
	title = {Discrete Higher Berry Phases and Matrix Product States},
	year = {2023}}

@article{Hsin_2023,
	author = {Po-Shen Hsin and Zhenghan Wang},
	doi = {10.1063/5.0136906},
	journal = {Journal of Mathematical Physics},
	month = {apr},
	number = {4},
	pages = {041901},
	publisher = {{AIP} Publishing},
	title = {On topology of the moduli space of gapped Hamiltonians for topological phases},
	url = {https://doi.org/10.1063%2F5.0136906},
	volume = {64},
	year = 2023}

@article{Aasen_2022,
	author = {David Aasen and Zhenghan Wang and Matthew B. Hastings},
	doi = {10.1103/physrevb.106.085122},
	journal = {Physical Review B},
	month = {aug},
	number = {8},
	publisher = {American Physical Society ({APS})},
	title = {Adiabatic paths of Hamiltonians, symmetries of topological order, and automorphism codes},
	url = {https://doi.org/10.1103%2Fphysrevb.106.085122},
	volume = {106},
	year = 2022}

@article{thouless_1983,
	author = {Thouless, D. J.},
	doi = {10.1103/PhysRevB.27.6083},
	issue = {10},
	journal = {Phys. Rev. B},
	month = {May},
	numpages = {0},
	pages = {6083--6087},
	publisher = {American Physical Society},
	title = {Quantization of particle transport},
	url = {https://link.aps.org/doi/10.1103/PhysRevB.27.6083},
	volume = {27},
	year = {1983}}

@article{Kapustin2201,
	adsurl = {https://ui.adsabs.harvard.edu/abs/2022JMP....63i1903K},
	archiveprefix = {arXiv},
	author = {{Kapustin}, Anton and {Sopenko}, Nikita},
	doi = {10.1063/5.0085964},
	eid = {091903},
	eprint = {2201.01327},
	journal = {Journal of Mathematical Physics},
	month = sep,
	number = {9},
	pages = {091903},
	primaryclass = {math-ph},
	title = {{Local Noether theorem for quantum lattice systems and topological invariants of gapped states}},
	volume = {63},
	year = 2022}

@article{2022aBachmann,
	adsurl = {https://ui.adsabs.harvard.edu/abs/2022arXiv220403763B},
	archiveprefix = {arXiv},
	author = {{Bachmann}, Sven and {De Roeck}, Wojciech and {Fraas}, Martin and {Jappens}, Tijl},
	doi = {10.48550/arXiv.2204.03763},
	eid = {arXiv:2204.03763},
	eprint = {2204.03763},
	journal = {arXiv e-prints},
	month = apr,
	pages = {arXiv:2204.03763},
	primaryclass = {math-ph},
	title = {{A classification of $G$-charge Thouless pumps in 1D invertible states}},
	year = 2022}

@unpublished{kitaevSimonsCenter1,
	author = {Kitaev, A.},
	date = {15 September 2011},
	note = {Talk at Simons Center for Geometry and Physics \url{http://scgp.stonybrook.edu/archives/1087}},
	title = {Toward a topological classification of many-body quantum states with short-range entanglement},
	year = {2011}}

@article{Manjunath_2025,
   title={Anomalous continuous symmetries and quantum topology of Goldstone modes},
   volume={111},
   ISSN={2469-9969},
   url={http://dx.doi.org/10.1103/PhysRevB.111.125151},
   DOI={10.1103/physrevb.111.125151},
   number={12},
   journal={Physical Review B},
   publisher={American Physical Society (APS)},
   author={Manjunath, Naren and Else, Dominic V.},
   year={2025},
   month=mar }

@misc{beaudry_2025,
      title={A Classifying Space for Phases of Matrix Product States}, 
      author={Agnes Beaudry and Michael Hermele and Markus J. Pflaum and Marvin Qi and Daniel D. Spiegel and David T. Stephen},
      year={2025},
      eprint={2501.14241},
      archivePrefix={arXiv},
      primaryClass={math-ph},
      url={https://arxiv.org/abs/2501.14241}, 
}

@unpublished{kitaevSimonsCenter2,
	author = {Kitaev, A.},
	date = {20 June 2013},
	note = {Talk at Simons Center for Geometry and Physics \url{http://scgp.stonybrook.edu/archives/16180}},
	title = {On the classification of short-range entangled states},
	year = {2013}}

@article{2023Ryu,
	adsurl = {https://ui.adsabs.harvard.edu/abs/2023arXiv230405356O},
	archiveprefix = {arXiv},
	author = {{Ohyama}, Shuhei and {Ryu}, Shinsei},
	doi = {10.48550/arXiv.2304.05356},
	eid = {arXiv:2304.05356},
	eprint = {2304.05356},
	journal = {arXiv e-prints},
	month = apr,
	pages = {arXiv:2304.05356},
	primaryclass = {cond-mat.str-el},
	title = {{Higher structures in matrix product states}},
	year = 2023}

@article{Kapustin2305,
	adsurl = {https://ui.adsabs.harvard.edu/abs/2023arXiv230506399A},
	archiveprefix = {arXiv},
	author = {{Artymowicz}, Adam and {Kapustin}, Anton and {Sopenko}, Nikita},
	eid = {arXiv:2305.06399},
	eprint = {2305.06399},
	journal = {arXiv e-prints},
	month = may,
	pages = {arXiv:2305.06399},
	primaryclass = {math-ph},
	title = {{Quantization of the higher Berry curvature and the higher Thouless pump}},
	year = 2023}

@unpublished{kitaev2019,
	author = {Kitaev, A.},
	date = {January 2019},
	note = {Talk at the conference in celebration of Dan Freed's 60th birthday: \url{https://web.ma.utexas.edu/topqft/talkslides/kitaev.pdf}},
	title = {Differential forms on the space of statistical mechanics models},
	year = {2019}}

@ARTICLE{2023Shiozaki,
       author = {{Shiozaki}, Ken and {Heinsdorf}, Niclas and {Ohyama}, Shuhei},
        title = "{Higher Berry curvature from matrix product states}",
      journal = {arXiv e-prints},
         year = 2023,
        month = may,
          eid = {arXiv:2305.08109},
        pages = {arXiv:2305.08109},
          doi = {10.48550/arXiv.2305.08109},
archivePrefix = {arXiv},
       eprint = {2305.08109},
 primaryClass = {quant-ph},
       adsurl = {https://ui.adsabs.harvard.edu/abs/2023arXiv230508109S}
}

@ARTICLE{2023Spodyneiko,
       author = {{Spodyneiko}, Lev},
        title = "{Hall conductivity pump}",
      journal = {arXiv e-prints},
         year = 2023,
        month = sep,
          eid = {arXiv:2309.14332},
        pages = {arXiv:2309.14332},
archivePrefix = {arXiv},
       eprint = {2309.14332},
 primaryClass = {cond-mat.mes-hall},
       adsurl = {https://ui.adsabs.harvard.edu/abs/2023arXiv230914332S}
}

@ARTICLE{2023Debray,
       author = {{Debray}, Arun and {Devalapurkar}, Sanath K. and {Krulewski}, Cameron and {Liu}, Yu Leon and {Pacheco-Tallaj}, Natalia and {Thorngren}, Ryan},
        title = "{A Long Exact Sequence in Symmetry Breaking: order parameter constraints, defect anomaly-matching, and higher Berry phases}",
      journal = {arXiv e-prints},
         year = 2023,
        month = sep,
          eid = {arXiv:2309.16749},
        pages = {arXiv:2309.16749},
          doi = {10.48550/arXiv.2309.16749},
archivePrefix = {arXiv},
       eprint = {2309.16749},
 primaryClass = {hep-th},
       adsurl = {https://ui.adsabs.harvard.edu/abs/2023arXiv230916749D}
}

@Article{2023_Qi,
	title={{Charting the space of ground states with tensor networks}},
	author={Marvin Qi and David T. Stephen and Xueda Wen and Daniel Spiegel and Markus J. Pflaum and Agnès Beaudry and Michael Hermele},
	journal={SciPost Phys.},
	volume={18},
	pages={168},
	year={2025},
	publisher={SciPost},
	doi={10.21468/SciPostPhys.18.5.168},
	url={https://scipost.org/10.21468/SciPostPhys.18.5.168},
}

@ARTICLE{2024_Sommer1,
       author = {{Sommer}, Ophelia Evelyn and {Wen}, Xueda and {Vishwanath}, Ashvin},
        title = "{Higher Berry Curvature from the Wave Function. I. Schmidt Decomposition and Matrix Product States}",
      journal = {\prl},
         year = 2025,
        month = apr,
       volume = {134},
       number = {14},
          eid = {146601},
        pages = {146601},
          doi = {10.1103/PhysRevLett.134.146601},
archivePrefix = {arXiv},
       eprint = {2405.05316},
 primaryClass = {cond-mat.str-el},
       adsurl = {https://ui.adsabs.harvard.edu/abs/2025PhRvL.134n6601S}
}

@ARTICLE{2024_Sommer2,
       author = {{Sommer}, Ophelia Evelyn and {Vishwanath}, Ashvin and {Wen}, Xueda},
        title = "{Higher Berry curvature from the wave function. II. Locally parametrized states beyond one dimension}",
      journal = {Phys. Rev. B},
         year = 2025,
        month = apr,
       volume = {111},
       number = {15},
          eid = {155110},
        pages = {155110},
          doi = {10.1103/PhysRevB.111.155110},
archivePrefix = {arXiv},
       eprint = {2405.05323},
 primaryClass = {cond-mat.str-el},
       adsurl = {https://ui.adsabs.harvard.edu/abs/2025PhRvB.111o5110S}
}

@ARTICLE{2024_Shuhei1,
       author = {{Ohyama}, Shuhei and {Ryu}, Shinsei},
        title = "{Higher Berry connection for matrix product states}",
      journal = {Phys. Rev. B},
         year = 2025,
        month = jan,
       volume = {111},
       number = {3},
          eid = {035121},
        pages = {035121},
          doi = {10.1103/PhysRevB.111.035121},
archivePrefix = {arXiv},
       eprint = {2405.05327},
 primaryClass = {cond-mat.str-el},
       adsurl = {https://ui.adsabs.harvard.edu/abs/2025PhRvB.111c5121O}
}

@ARTICLE{2024_Shuhei2,
       author = {{Ohyama}, Shuhei and {Ryu}, Shinsei},
        title = "{Higher Berry phase from projected entangled pair states in (2+1) dimensions}",
      journal = {Phys. Rev. B},
         year = 2025,
        month = jan,
       volume = {111},
       number = {4},
          eid = {045112},
        pages = {045112},
          doi = {10.1103/PhysRevB.111.045112},
archivePrefix = {arXiv},
       eprint = {2405.05325},
 primaryClass = {cond-mat.str-el},
       adsurl = {https://ui.adsabs.harvard.edu/abs/2025PhRvB.111d5112O}
}

@book{2023spectral,
  title={Spectral flow: A functional analytic and index-theoretic approach},
  author={Doll, Nora and Schulz-Baldes, Hermann and Waterstraat, Nils},
  year={2023},
  publisher={De Gruyter}
}

@misc{wen_2025,
      title={Space of conformal boundary conditions from the view of higher Berry phase: Flow of Berry curvature in parametrized BCFTs}, 
      author={Xueda Wen},
      year={2025},
      eprint={2507.12546},
      archivePrefix={arXiv},
      primaryClass={hep-th},
      url={https://arxiv.org/abs/2507.12546}, 
}

@misc{choi_2025,
      title={Higher Structures on Boundary Conformal Manifolds: Higher Berry Phase and Boundary Conformal Field Theory}, 
      author={Yichul Choi and Hyunsoo Ha and Dongyeob Kim and Yuya Kusuki and Shuhei Ohyama and Shinsei Ryu},
      year={2025},
      eprint={2507.12525},
      archivePrefix={arXiv},
      primaryClass={hep-th},
      url={https://arxiv.org/abs/2507.12525}, 
}

@misc{Shiozaki_2025,
      title={Equivariant Parameter Families of Spin Chains: A Discrete MPS Formulation}, 
      author={Ken Shiozaki},
      year={2025},
      eprint={2507.19932},
      archivePrefix={arXiv},
      primaryClass={quant-ph},
      url={https://arxiv.org/abs/2507.19932}, 
}

@article{Fulga_2012,
   title={Scattering theory of topological insulators and superconductors},
   volume={85},
   ISSN={1550-235X},
   url={http://dx.doi.org/10.1103/PhysRevB.85.165409},
   DOI={10.1103/physrevb.85.165409},
   number={16},
   journal={Physical Review B},
   publisher={American Physical Society (APS)},
   author={Fulga, I. C. and Hassler, F. and Akhmerov, A. R.},
   year={2012},
   month=apr }

@article{Brouwer_1998,
   title={Scattering approach to parametric pumping},
   volume={58},
   ISSN={1095-3795},
   url={http://dx.doi.org/10.1103/PhysRevB.58.R10135},
   DOI={10.1103/physrevb.58.r10135},
   number={16},
   journal={Physical Review B},
   publisher={American Physical Society (APS)},
   author={Brouwer, P. W.},
   year={1998},
}

@article{Br_unlich_2009,
   title={Equivalence of Topological and Scattering Approaches to Quantum Pumping},
   volume={295},
   ISSN={1432-0916},
   url={http://dx.doi.org/10.1007/s00220-009-0983-1},
   DOI={10.1007/s00220-009-0983-1},
   number={1},
   journal={Communications in Mathematical Physics},
   publisher={Springer Science and Business Media LLC},
   author={Bräunlich, G. and Graf, G. M. and Ortelli, G.},
   year={2009},
   month=dec
}

@article{PhysRevB.83.155429,
  title = {Scattering formula for the topological quantum number of a disordered multimode wire},
  author = {Fulga, I. C. and Hassler, F. and Akhmerov, A. R. and Beenakker, C. W. J.},
  journal = {Phys. Rev. B},
  volume = {83},
  issue = {15},
  pages = {155429},
  numpages = {8},
  year = {2011},
  month = {Apr},
  publisher = {American Physical Society},
  doi = {10.1103/PhysRevB.83.155429},
  url = {https://link.aps.org/doi/10.1103/PhysRevB.83.155429}
}

@article{Schulz_Baldes_2020,
   title={Dimensional Reduction and Scattering Formulation for Even Topological Invariants},
   volume={381},
   ISSN={1432-0916},
   url={http://dx.doi.org/10.1007/s00220-020-03886-y},
   DOI={10.1007/s00220-020-03886-y},
   number={1},
   journal={Communications in Mathematical Physics},
   publisher={Springer Science and Business Media LLC},
   author={Schulz-Baldes, Hermann and Toniolo, Daniele},
   year={2020},
   month=nov
}

\end{document}